\documentclass[11pt,a4paper]{amsart}

\usepackage{natbib}

\usepackage{hyperref}
\hypersetup{
    colorlinks=true,
    linkcolor=blue,
    citecolor=red,
    pdfpagemode=FullScreen,
    pdftitle={Aletti,Crimaldi,Ghiglietti}
    }

\usepackage[a4paper, total={6.8in, 9in}]{geometry}

\theoremstyle{plain}
\newtheorem{thm}{Theorem}[section]
\newtheorem{lem}[thm]{Lemma}

\newtheorem{cor}[thm]{Corollary}
\newtheorem{rem}{Remark}[section]
\theoremstyle{definition}

\newtheorem{asmp}{Assumption}[section]

\numberwithin{equation}{section} 

\allowdisplaybreaks[4]

\newcommand{\veta}{{\boldsymbol{\eta}}}
\usepackage{tikz}
\foreach \a in {q,w,e,r,t,y,u,i,o,p,a,s,d,f,g,h,j,k,l,z,x,c,v,b,n,m,Q,W,E,R,T,Y,U,I,O,P,A,S,D,F,G,H,J,K,L,Z,X,C,V,B,N,M}{
  \expandafter\xdef\csname v\a\endcsname {
		{\noexpand\mathbf{\a}}
	}
}

\newcommand{\vone}{{\boldsymbol{1}}}
\newcommand{\vzero}{{\boldsymbol{0}}}

\begin{document}

\title[Central limit theorems for interacting innovation processes]{Central limit theorems for interacting innovation processes, related statistical tools and general results}

\author[G. Aletti]{Giacomo Aletti}
\address{ADAMSS Center,
  Universit\`a degli Studi di Milano, Milan, Italy}
\email{giacomo.aletti@unimi.it}

\author[I. Crimaldi]{Irene Crimaldi}
\address{IMT School for Advanced Studies Lucca, Lucca, Italy}
\email{irene.crimaldi@imtlucca.it}

\author[A. Ghiglietti]{Andrea Ghiglietti}
\address{Universit\`a degli Studi di Milano-Bicocca, Milan, Italy}
\email{andrea.ghiglietti@unimib.it (Corresponding author)}

\thanks{G.\ Aletti is a member of ``Gruppo Nazionale per il Calcolo
  Scientifico (GNCS)'' of the Italian Institute ``Istituto Nazionale
  di Alta Matematica (INdAM)''. I.\ Crimaldi and A.\ Ghiglietti are members of ``Gruppo
  Nazionale per l'Analisi Matematica, la Probabilit\`a e le loro
  Applicazioni (GNAMPA)'' of the Italian Institute ``Istituto
  Nazionale di Alta Matematica (INdAM)''. Irene Crimaldi thanks the “Resilienza Economica e Digitale” (RED) 
project (CUP
D67G23000060001) funded by the Italian Ministry of University and Research (MUR)
as “Department of Excellence” (Dipartimenti di Eccellenza 2023-2027, Ministerial
Decree no. 230/2022). Partially supported by the European Union - NextGenerationEU through the Italian Ministry of University
and Research under the National Recovery and Resilience Plan (PNRR) - M4C2 - Investment 1.3, 
 title [Public Sector Indicators for Sustainability and Wellbeing (PUWELL)] - 
 Program [Growing Resilient, INclusive and Sustainable (GRINS)] - PE18 - CUP J33C22002910001.}

\begin{abstract}
We study a networked system of innovation processes, where each process is modeled as an urn with infinitely many colors-a classical framework for capturing the emergence of novelties. Extending this paradigm, we analyze a model of interacting urns, where the probability of generating or reusing elements in one process is influenced by the histories of others. This interaction is governed by two matrices that control innovation triggering and reinforcement dynamics across the system. The core contribution of this work is a detailed analysis of the second-order asymptotic behavior of the model. Building on these theoretical results, we develop statistical tools to infer the structure and strength of inter-process influence. The methodology is framed in a general setting, making it broadly applicable. We validate our approach with applications to two real-world datasets from Reddit discussions and Gutenberg text corpora.

\end{abstract}

\maketitle

\section{Introduction}

Understanding how novelties emerge, propagate, and catalyze further innovations is crucial in several disciplines, including biology, linguistics, and social sciences \cite{armano, arthur, fink, puglisi, rogers, rzh}. In probabilistic terms, a \emph{novelty} (or \emph{innovation}) is the first occurrence of an event of interest.\\

\indent Widely adopted mathematical models for such \emph{innovation processes} include \emph{urn models with infinitely many colors} \cite{ban-tha-2022, Mailler20, Sariev23, Sariev24} and \emph{species sampling sequences} \cite{favaro09, han-pitman, lijoi07, lijoi2010, pitman_1996}. In the urn metaphor, for each $t\in{\mathbb N}\setminus \{0\}$ we denote by $C_t$ the observed color at extraction $t$. The probability that the $(t+1)$-th draw yields a ``new'' color (i.e., one not previously observed) is $Z^*_t$, a function of $C_1,\dots,C_t$, while the probability of drawing an ``old'' (i.e., already observed) color $c$ is
$P_{c,t}=\sum_{n=1}^t Q_{n,t}\,\mathbb{I}_{\{C_n=c\}}$,
where $Q_{n,t}$ depends on $C_1,\dots,C_t$. The quantities $Z^*_t$ and $Q_{n,t}$ specify the model. In this framework, the set of possible colors is unbounded, so new colors may continuously enter the system: the urn represents the space of possibilities, and the extracted sequence the realized history.

\indent A canonical example is the \emph{Poisson-Dirichlet (PD) model} \cite{james08, pit-yor}. In this model,
\begin{equation}\label{PD-probab}
  \begin{split}
  &Z^*_t=\frac{\theta+\gamma D_t}{\theta +  t},
  \qquad Q_{n,t}=\frac{1-\gamma/K_{C_n,t}}{\theta + t},
  \\
&\mbox{and so }\quad
P_{c,t}=
\frac{K_{c,t}-\gamma}{\theta +  t},
\end{split}
\end{equation}
where $0\leq \gamma<1$, $\theta>-\gamma$, $D_t$ is the number of distinct colors observed up to time-step $t$, and $K_{c,t}$ is the count of color $c$ up to time-step $t$. This model embodies the ``preferential attachment'' (or ``popularity'') principle, since the probability of observing a color is proportional to its frequency. The PD process effectively captures empirical laws such as \emph{Heaps' law} \citep{heaps_1978}, describing the sublinear growth of distinct elements with the number of observations, and \emph{Zipf's law} \citep{zipf_1949}, characterizing the sublinear decrease of ranked frequencies. In particular, in the PD model $D_t\propto t^\gamma$ for $0<\gamma<1$, and $D_t\propto \ln(t)$ when $\gamma=0$ \cite{janson-2005,Tria1}.\\

\indent An intuitive representation of the PD process is the \emph{urn with triggering} model \cite{ale-cri_SR, Tria2} (or the equivalent formulation in \cite{Zabell_2005}). Initially, the urn contains $N_0>0$ distinct balls of different colors. At each time-step $t+1$, a ball is drawn at random with replacement and:
\begin{itemize}
\item if the drawn color is \emph{new}, we add $(\widehat{\rho}-1)$ balls of the same color and $(\nu+1)$ balls of entirely new (distinct) colors;
\item if the drawn color is \emph{old}, we add $\rho$ balls of the same color.
\end{itemize}
When the \emph{balance condition} $\widehat{\rho}+\nu=\rho$ holds (so that the number of added balls is constant), the model reproduces \eqref{PD-probab} by taking $\rho>\nu\geq 0$ and setting $\theta=N_0/\rho$ and $\gamma=\nu/\rho$.\\

\indent This formulation highlights both \emph{reinforcement} and \emph{Kauffman's principle of the adjacent possible} \citep{kauffman_2000}: when a novelty occurs, it expands the space of potential future novelties by introducing new colors.

\indent To study how distinct innovation processes influence each other, in \cite{ale-cri-ghi-innovation-2023} we introduced and analyzed a finite network of urns with triggering (see Section~\ref{model}). Each node has its own urn, and the probability of observing a new or old item at a node depends on both the node's history and the histories of the other nodes. Interactions are governed by two matrices: $\Gamma$ and $W$. Matrix $\Gamma$ modulates how novelties produced in the system affect each node's probability of generating new novelties, extending the adjacent possible to the system level. Matrix $W$ regulates reinforcement across nodes, making the probability of extracting already observed items depend on their frequencies in all nodes.\\

\indent When $\Gamma$ and $W$ are irreducible, the model captures synchronization: all nodes produce innovations at the same asymptotic rate, and the distribution of observed items becomes uniform across the network. These results are formalized in Theorems~\ref{th-synchro-rates} and~\ref{th-synchro-probab}.\\

\indent This work aims to:
\begin{itemize}
    \item derive second-order asymptotic behaviour of the model by providing Central Limit Theorems (CLTs) for the quantities of interest;
    \item develop statistical tools for inference on the interaction matrices $\Gamma$ and $W$, as well as on the limiting probabilities of observing an extracted item $c$;
    \item apply the theoretical findings to real-world datasets, assessing interaction intensities in empirical contexts.
\end{itemize}

\indent We also prove general theoretical results that apply beyond the present setting. They concern a pair of multidimensional stochastic processes following a suitable recursive dynamics. Since the asymptotic covariance matrices in our results are random, we do not rely on the standard convergence in distribution, but instead we employ the notion of \emph{stable convergence}.

\subsection{Related literature}
Although interacting urn models with a finite set of colors are widely studied (see, e.g., \cite{ale-cri-ghi, ale-cri-ghi-complete} and references therein), works on urns with infinitely many colors are comparatively scarce. Notable exceptions are \citep{fortini, iacopini-2020}, which investigate Dirichlet processes with random reinforcement and urns with triggering, respectively. These approaches differ from ours in the definition of novelty and in the interaction structure, in particular concerning the system-level adjacent possible and the dual reinforcement mechanism.

In \cite{fortini} (see Example~3.8 therein), a finite collection of Dirichlet processes with random reinforcement is considered. A random weight $W_{t,h}$ is associated with the extraction at time-step $t$ from urn $h$, and the probability of extracting an old color (where ``old'' is defined at the level of urn $h$ only) is proportional to the weight of that color. Interactions across urns are introduced through stochastic dependence among the weights: for instance, the weights may be shared across urns, depend on outcomes of other urns, or depend on common observable or latent variables.

The present work and \cite{iacopini-2020} both study interacting urns with triggering, where interaction acts when a new color is extracted, thus extending Kauffman's principle from a single agent to a network. In contrast with \cite{iacopini-2020}, in our model the extraction of an ``old'' item is also interactive: the probability of extracting an old item depends on its total number of occurrences across all nodes. Moreover, \cite{iacopini-2020} focuses on node-level novelties (``discoveries''), i.e.\ first appearances in a given node, possibly already observed in other nodes, whereas we study the sequence of novelties for the whole network produced by each node.

\subsection{Structure of the work}
The remainder of the paper is organized as follows. In Section~\ref{model} we describe the model and recall the first-order convergence results. In Section~\ref{main-CLTs} we present the main results, consisting of CLTs describing the second-order asymptotic behaviour of the interacting innovation processes. Then, in Section~\ref{stat-inference} we illustrate statistical tools derived from the CLTs. The inferential methodology is presented for a suitable bi-dimensional recursive process in a separate appendix, including theoretical results and proofs; sections, theorems, and remarks in the appendix are referenced with capital letters. Finally, in Section~\ref{experiments} we apply the proposed methods to two real datasets, evaluating interaction intensities.


\section{Model and first-order convergence results}\label{model}

Suppose we have $N$ urns, labeled from $1$ to $N$.  
{At time-step $0$, the colors inside each urn are different from those in the other urns}. 
Let $N_{0,h}>0$ be the number of distinct balls with distinct colors inside the urn $h$.   
Then, at each time-step $t\geq 1$, one ball is randomly drawn with replacement
from each urn and, for any $h=1,\dots N$, urn $h$ is so updated according to the colors extracted
 from urn $h$ itself and from all the other urns $j\neq h$:
\begin{itemize}
\item if the color $C_{t,h}$ of the ball extracted from urn $h$ is ``new'' (i.e., it appears
  for the first time in the system), then into urn $h$ we add    
  $(\widehat{\rho}_{h,h}-1)$, with $\widehat{\rho}_{h,h}>0$, balls of the same color as the extracted one 
  plus $(\nu_{h,h}+1)$, with $\nu_{h,h}\geq 0$, 
  balls of distinct ``new'' colors (i.e. not already present in the system);
 
 \item if the color $C_{t,h}$ of the ball extracted from urn $h$ is ``old'' (i.e., it has
   been already extracted in the system), we add $\rho_{h,h}>0$ balls of the same color 
   into urn $h$; 

\item for each $j\neq h$, if the color $C_{t,j}$ of the ball extracted from urn $j$ is ``new'' (i.e., it appears
  for the first time in the system), then into urn $h$ we add $\widehat{\rho}_{j,h}\geq 0$ balls of 
  the same color as the one extracted from urn $j$ 
  plus $\nu_{j,h}\geq 0$ balls of distinct ``new'' colors (i.e. not already present in the system);
  
 \item for each $j\neq h$, if the color $C_{t,j}$ of the ball extracted from urn $j$ is ``old'' (i.e., it has
   been already extracted in the system), then into urn $h$ we add $\rho_{j,h}\geq 0$ 
   balls of the same color as the one extracted from urn $j$.
 \end{itemize}
The terms ``new'' and ``old'' refer to the entire system. We assume that the ``new'' colors added to a given urn are always different from those added to the other urns (at the same time-step or in the past). Together with the initial condition, this implies that the same novelty cannot be observed simultaneously in different urns. Hence, for each observed novelty $c$, there exists a unique urn, denoted $j^*(c)$, that produced it. After its first extraction, $c$ may subsequently appear in other urns due to interactions among urns.
Furthermore, as in the standard Poisson-Dirichlet model, we assume the {\em balance condition} 
\begin{equation}\label{balance-cond}
\widehat{\rho}_{j,h}+\nu_{j,h}=\rho_{j,h},\quad \mbox{i.e. } \widehat{\rho}_{j,h}=\rho_{j,h}-\nu_{j,h}\,,
\end{equation} 
so that, at each time-step, each urn $j$ contributes to increase the number of balls inside urn $h$ by $\rho_{j,h}\geq 0$, 
with $\rho_{h,h}>0$. Moreover,
letting $\rho_h=\sum_{j=1}^N \rho_{j,h}$ 
and  setting (without loss of generality) 
\begin{equation}\label{normalization}
\theta_h=N_{0,h}/\rho_h,\quad \gamma_{j,h}=\nu_{j,h}/\rho_h,\quad  
\lambda_{j,h}=\widehat{\rho}_{j,h}/\rho_h\quad\mbox{and}\quad 
w_{j,h}=\rho_{j,h}/\rho_h\,, 
\end{equation}
we obtain
\begin{equation}\label{birth-prob-inter}
Z^*_{t,h}=P(C_{t+1,h}=\mbox{``new''}\,|\, \vC_1,\dots,\vC_t)=
\frac{\theta_{h}+\sum_{j=1}^N \gamma_{j,h}D^*_{t,j}}{\theta_{h} + t}\,,
\end{equation}
where $\vC_n=(C_{n,1},\dots,C_{n,N})^\top$ with $n=1,\dots,t$,
$D^*_{t,j}$ denotes the number, until time-step $t$, 
of distinct observed colors extracted for their first time from urn $j$, 
 that is the number of distinct novelties for the whole system 
 ``produced'' by urn (process) $j$ until time-step $t$, 
and, for each ``old'' color $c$, 
\begin{equation}\label{old-color-prob-inter}
\begin{split}
P_{t}(h,c)&=P(C_{t+1,h}= c \,|\, \vC_1,\dots,\vC_t )\\
&= 
\frac{\sum_{j\neq j^*(c)}w_{j,h}K_t(j,c)+w_{j^*(c),h}(K_t(j^*(c),c)-1)+\lambda_{j^*(c),h}}{\theta_h + t}
\\
&=\frac{\sum_{j=1}^N w_{j,h}K_t(j,c)-\gamma_{j^*(c),h}}{\theta_h + t}\,, 
\end{split}
\end{equation}
where $K_{t}(j,c)$ denotes the number of times the color $c$ has been extracted from urn $j$ until time-step $t$ and 
$j^*(c)$ denotes the urn from which the color $c$ has been extracted for the first time.
Note that the probability that urn $h$ produces at time-step $t+1$ a novelty for the entire system depends increasingly on the number $D^*_{t,j}$ of novelties produced by urn $j$ up to time $t$, and $\gamma_{j,h}$ regulates this dependence. Thus, Kauffman's principle of the adjacent possible holds at the system level: for each pair $(j,h)$, $\gamma_{j,h}$ quantifies how the production of a novelty by urn $j$ induces potential novelties in urn $h$. Similarly, the probability that urn $h$ extracts an old color $c$ at time-step $t+1$ depends increasingly on the counts $K_t(j,c)$, with $w_{j,h}$ quantifying the influence of urn $j$ on urn $h$ through reinforcement.

In \cite{ale-cri-ghi-innovation-2023} (see also~\cite[Appendix~B.1]{ale-cri-ghi-nut} for a general theory) 
we proved the first-order asymptotic properties of the model:

\begin{thm}[{\cite[Theorem~3.1]{ale-cri-ghi-innovation-2023}}]\label{th-synchro-rates}
Suppose that the matrix $\Gamma=(\gamma_{j,h})_{j,h=1,\dots,N}$ is irreducible. Denote by $\gamma^*\in (0,1)$ the 
Perron-Frobenius eigenvalue of $\Gamma$ and by $\vu$ its 
left eigenvector. 
Then, we have 
\begin{equation*}\label{as-Dstar}
t^{-\gamma^*}\vD^*_{t}\stackrel{a.s.}\longrightarrow D^{**}_{\infty}\vu \,,
\end{equation*}
where $\vD^*_t = (D^*_{t,1},\ldots, D^*_{t,N})^\top$ and
 $D^{**}_{\infty}$ is an integrable strictly positive random variable.
\end{thm}

\begin{thm}[{\cite[Theorem~3.2]{ale-cri-ghi-innovation-2023}}]\label{th-synchro-probab}
Suppose that the matrix $W=(w_{j,h})_{j,h=1,\dots,N}$ is irreducible.
Then, for 
each observed color $c$ in the system,
we have
\begin{equation*}\label{as-P}
\vP_t(c)\stackrel{a.s.}\longrightarrow \widetilde{P}_\infty(c)\vone
\end{equation*}
\begin{equation*}\label{as-Kmedio}
 \frac{1}{t}\vK_t(c)\stackrel{a.s.}\longrightarrow \widetilde{P}_{\infty}(c)\vone
  \quad\mbox{and so}\quad
  \frac{K_t(h,c)}{\sum_{j=1}^N K_t(j,c)} \stackrel{a.s.}\longrightarrow \frac{1}{N}\,,
\end{equation*}
where $\vP_t(c)=(P_{t}(1,c),\dots,P_{t}(N,c))^\top$, $\vK_t(c)=(K_{t}(1,c),\dots,K_{t}(N,c))^\top$,  
${\vone}$ denotes the vector with all the components equal to~$1$ 
and 
$\widetilde{P}_{\infty}(c)$ is a random variable that
takes values in $(0,1)$.
\footnote{Note that in \cite[Theorem~3.2 ]{ale-cri-ghi-innovation-2023}
we state that this random variable takes values in $(0,1]$, but, 
applying \cite[Theorem~S1.3]{ale-cri-ghi-innovation-2023}
also replacing $P_t(h,c)$ by $1-P_t(h,c)$, we immediately obtain
$(1-\widetilde{P}_\infty(c))>0$ a.s., that is $\widetilde{P}_\infty(c)<1$ a.s..}
\end{thm}

In the next section we complete the analysis of the asymptotics of the  model providing some 
second-order convergence results.


\section{Central limit theorems}\label{main-CLTs}

 Assume $N\geq 2$. Let $\Gamma$, $W$, $\Lambda$ be equal to
 the {\em non-negative} $N\times N$ square matrices 
 with elements equal to the model parameters
 $\gamma_{j,h}$, $w_{j,h}$ (with $w_{h,h}>0$) and 
 $\lambda_{j,h}$ (with $\lambda_{h,h}>0$), respectively.
 We recall that, by the balance condition~\eqref{balance-cond}  
and the reparametrization \eqref{normalization}, we have 
\begin{equation}\label{ass:interaction_matrices}
    W=\Gamma+\Lambda,\qquad
    {\vone}^\top\Gamma<\vone^\top
    \qquad \mbox{and}\qquad {\vone}^\top W={\vone}^\top,
\end{equation}
where ${\vone}$ denotes the vector with all the components equal to~$1$.
As observed above, the matrix $\Gamma$ rules the production of potential novelties and, in particular, 
its elements out of the diagonal regulate the interaction among the innovation
processes with respect to this issue; while, 
the matrix $W$ rules the interaction among the innovation processes with respect to the choice of an old item. 
Moreover, recall that 
$\vD^*_t=(D^*_{t,1},\dots,D^*_{t, N})^\top$,   
$\vP_t(c)=(P_{t}(1,c),\dots,P_{t}(N,c))^\top$ and $\vK_t(c)=(K_{t}(1,c),\dots,K_{t}(N,c))^\top$. 
Finally, the symbol ${\vzero}$ will denote the vector with all the components equal to~$0$.
\\

\indent In order to simplifying the statements of the following theorems, we summarize the required notation and conditions
 in the following two sets of assumptions:
 
\begin{asmp}\label{ass-Gamma}
Assume that the non-negative matrix $\Gamma=(\gamma_{j,h})_{j,h=1,\dots,N}$ is irreducible and diagonalizable. Denote by $\gamma^*\in (0,1)$ the 
Perron-Frobenius eigenvalue of $\Gamma$, by ${\vv}$ the corresponding right eigenvector 
with strictly positive entries and 
such that ${\vv}^\top{\vone}=1$ and  denote 
by ${\vu}$ the corresponding left eigenvector  
with strictly positive entries and  such that 
${\vv}^\top{\vu}=1$. Finally, let $\gamma^*_2$ be an
  eigenvalue of $\Gamma$ different from $\gamma^*$ and with highest real part  
and assume that ${\mathcal Re}(\gamma^*_2)/\gamma^*<1/2$. 
\end{asmp}

\begin{asmp}\label{ass-W}
  Assume that the non-negative matrix $W=(w_{j,h})_{j,h=1,\dots,N}$ is irreducible and diagonalizable. Denote by ${\vv_W}$ the right eigenvector associated to the leading
  eigenvalue $w^*=1$,  
  with strictly positive entries and
  such that ${\vv_W}^\top{\vone}=1$. Finally, let
  $w^*_2$ be an eigenvalue of $W$ different from $w^*=1$ and
    with highest real part  
and assume that ${\mathcal Re}(w^*_2)<\frac{1}{2}$. 
  \end{asmp}

We are now ready to provide the central limit theorems for the considered
model, that describe the second-order behaviour of the quantities in Theorems~\ref{th-synchro-rates} and~\ref{th-synchro-probab} under the above assumptions. 
A discussion of possible extensions of these results is postponed to  Remark~\ref{rem:CLT when assumption second eigenvalue}, after the proofs, at the end of this section. For completeness, we also point out that the special case $N=1$ is investigated in \cite{Bercu24, Gouet93}.

 \begin{thm}\label{CLT-Dstar}  
Under Assumption \ref{ass-Gamma},  we have 
$$
t^{\gamma^*/2}\left(\frac{\vD_t^*}{t^{\gamma^*}}-D^{**}_\infty\vu\right)
\stackrel{stably}\longrightarrow 
\mathcal{N}(\vzero, D^{**}_\infty \gamma^* C_{det,\Gamma})\,,
$$
where $D^{**}_\infty$ is defined in Theorem~\ref{th-synchro-rates}
and
$C_{det,\Gamma}$ is a deterministic matrix depending on the
eigen-structure of $\Gamma$, which is possible to
write explicitly.
 \end{thm}
 \begin{proof} Denote by $X^*_{t,h}$ the random variable that takes value $1$ when the ball extracted from urn $h$ at time-step $t$ has 
a new (for all the system) color and is equal to $0$ otherwise. Then $Z^*_{t,h}$ 
coincides with $P(X^*_{t+1,h}=1\,|\,\vC_1,\dots,\vC_t)=
E[X^*_{t+1,h}\,|\,\vC_1,\dots,\vC_t]$ and 
$D^*_{t,h}$ can be written as $\sum_{n=1}^t X^*_{n,h}$. Since we have  
 $$
 Z^*_{t,h}=\frac{\theta_h+\sum_{n=1}^t\sum_{j=1}^N \gamma_{j,h}X^*_{n,j}}{\theta_h+t}\,,
 $$
 we obtain the following dynamics for $Z^*_{t,h}$:
 \begin{equation*}
 Z^*_{0,h}=1,\qquad Z^*_{t+1,h}=(1-r_{t,h})Z^*_{t,h}+r_{t,h}\sum_{j=1}^N \gamma_{j,h}X^*_{t+1,j}\quad\mbox{for } t\geq 0\,,
 \end{equation*}
 where $r_{t,h}=1/(\theta_h+t+1)=1/(t+1)+O(1/t^2)$
 \footnote{In all the paper, the notation $O(s_t)$ denotes a generic (possibly random) remainder term $R_t$
   such that $|R_t|\leq C s_t$ for a suitable deterministic constant $C$ and for $t$ large enough.}. 
   The corresponding vectorial dynamics
 for $\vZ^*_t=(Z^*_{t,1},\dots,Z^*_{t,N})^\top$ is
\begin{equation}\label{eq-dynamics-vector}
  \begin{split}
    {\vZ}^*_0&=\vone\\
    {\vZ}^*_{t+1}&=
    \left(1-\frac{1}{t+1}\right){\vZ}^*_{t}+\frac{1}{t+1} \Gamma^\top {\vX}_{t+1}^* + O(1/t^2)\vone\\
    &=
    \left[1-\frac{1}{t+1}(I-\Gamma^\top)\right]{\vZ}^{*}_{t}+
    \frac{1}{t+1}\Gamma^\top\Delta {\vM}^{*}_{t+1} +
    O(1/t^2)\vone
\quad\mbox{for } t\geq 0,
\end{split}
\end{equation}
where $\Delta{\vM}^*_{t+1}={\vX}^*_{t+1}-{\vZ}^*_t$. Now, we fix $x>0$ and set
$$
\zeta_0(x)=1,\qquad \zeta_{t}(x)=
\frac{\Gamma(t+x)}{\Gamma(t)}\sim
t^{x}\uparrow +\infty \,.
$$
More precisely, from \cite[Lemma 4.1]{Gouet93} we have
\begin{equation*}
\zeta_t(x)=t^x + O(t^{x-1})\quad\mbox{and}\quad \frac{1}{\zeta_t(x)}=\frac{1}{t^x}+O(1/t^{x+1}),
\end{equation*}
and so 
\begin{equation}\label{eq-ordine}
\begin{split}
\frac{1}{\zeta_{t+1}(x)\zeta_{t+1}(1-x)}&= 
\left( \frac{1}{(t+1)^x}+O(1/t^{x+1}) \right) \left( \frac{1}{(t+1)^{1-x}}+O(1/t^{2-x}) \right)\\ 
&=
\frac{1}{t+1}+O(1/t^2).
\end{split}
\end{equation}

\indent Hence,  multiplying \eqref{eq-dynamics-vector} by $\zeta_{t+1}(1-\gamma^*)$ and using the relation 
$$
\frac{\zeta_{t+1}(x)}{\zeta_t(x)}=
\frac{\Gamma(t+x+1)}{\Gamma(t+1)}
\frac{\Gamma(t)}{\Gamma(t+x)}=
1+\frac{x}{t}=
1+\frac{x}{t+1}+O(1/t^2),$$
with $x=1-\gamma^*$, 
 we get the following dynamics for ${\vZ}^{**}_{t}=\zeta_{t}(1-\gamma^*){\vZ}^*_{t}$, where we set 
 $\Delta{\vM}^{**}_{t+1}=\zeta_t(1-\gamma^*)\Delta{\vM}^*_{t+1}$: 
 \begin{equation*}
\begin{split}
{\vZ}^{**}_0&=\vone\\ 
{\vZ}^{**}_{t+1}&=
\left[1-\frac{1}{t+1}(I-\Gamma^\top)\right]\frac{\zeta_{t+1}(1-\gamma^*)}{\zeta_t(1-\gamma^*)}{\vZ}^{**}_{t}+
\frac{1}{t+1}\Gamma^\top\Delta {\vM}^{**}_{t+1}+O\left(\frac{\zeta_{t+1}(1-\gamma^*)}{t^2}\right)\vone
\\
&=
\left[1-\frac{1}{t+1}(I-\Gamma^\top)\right]
\left(1+\frac{1-\gamma^*}{t+1}+O(1/t^2)\right){\vZ}^{**}_{t}+
\frac{1}{t+1}\Gamma^\top\Delta {\vM}^{**}_{t+1}+O\left(\frac{\zeta_{t+1}(1-\gamma^*)}{t^2}\right)\vone
\\
&=
{\vZ}^{**}_{t}-\frac{1}{t+1}(\gamma^* I-\Gamma^\top){\vZ}^{**}_{t}+
\frac{1}{t+1}\Gamma^\top\Delta {\vM}^{**}_{t+1}+O\left(\frac{\zeta_{t+1}(1-\gamma^*)}{t^2}\right)\vone\,.
\end{split}
\end{equation*}

Moreover, recalling that $D^*_{t,h}=\sum_{n=1}^tX^*_{n,h}$, we find
\begin{equation*}
\begin{split}
\frac{\vD^*_0}{\zeta_0(\gamma^*)}&=\vzero\\
\frac{\vD^*_{t+1}}{\zeta_{t+1}(\gamma^*)}&=
\frac{\zeta_t(\gamma^*)}{\zeta_{t+1}(\gamma^*)}\frac{\vD_t^*}{\zeta_t(\gamma^*)}+
\frac{1}{\zeta_{t+1}(\gamma^*)}\vX^*_{t+1}\\
&=
\left(1-\frac{\gamma^*}{t+1} + O(t^{-2})\right) \frac{\vD_t^*}{\zeta_t(\gamma^*)}+
\frac{1}{\zeta_{t+1}(\gamma^*)}\Delta \vM^*_{t+1}+\frac{1}{\zeta_{t+1}(\gamma^*)}\vZ^*_t\\
&=\left( 1-\frac{\gamma^*}{t+1} + O(t^{-2})\right) \frac{\vD_t^*}{\zeta_t(\gamma^*)}+
\frac{1}{\zeta_{t+1}(\gamma^*)\zeta_{t+1}(1-\gamma^*)} \Delta\vM^{**}_{t+1}\\
&+ \frac{1}{\zeta_{t+1}(\gamma^*)\zeta_{t+1}(1-\gamma^*)}\frac{\zeta_{t+1}(1-\gamma^*)}{\zeta_t(1-\gamma^*)} \vZ^{**}_t\,.
\end{split}
\end{equation*}

Using \eqref{eq-ordine} and the relation $\zeta_{t+1}(x)/\zeta_t(x)=1+O(1/t)$, 
and combining together with the previous dynamics of $({\vZ}^{**}_{t})_t$ and $(\vD^*_{t+1}/\zeta_{t+1}(\gamma^*))_t$, we obtain 
\begin{equation*}\label{eq-system-1}
\begin{split}
{\vZ}^{**}_{t+1}&={\vZ}^{**}_{t}-\frac{1}{t+1}(\gamma^* I-\Gamma^\top){\vZ}^{**}_{t}+
\frac{1}{t+1}\Gamma^\top\Delta {\vM}^{**}_{t+1}+O\left(\zeta_{t+1}(1-\gamma^*)/t^2\right)\vone
\\
\frac{\vD^*_{t+1}}{\zeta_{t+1}(\gamma^*)}&=
\frac{\vD_t^*}{\zeta_t(\gamma^*)} - \frac{1}{t+1}\left(\gamma^*\frac{\vD_t^*}{\zeta_t(\gamma^*)}-\vZ^{**}_t \right) +
\frac{1}{t+1} \Delta\vM^{**}_{t+1} + O(\zeta_t(1-\gamma^*)/t^2)\vone\,,
\end{split}
\end{equation*}
where $ O(\zeta_t(1-\gamma^*)/t^2)=O(1/t^{1+\gamma^*})$. 
Moreover, we recall that in \cite[Theorem S1.1 and Lemma S1.2]{ale-cri-ghi-innovation-2023} 
 (see also [8, Appendix B.1] for a general theory)
we proved
\begin{equation*}
  \begin{split}
    \widetilde{Z}^{**}_{t}=\zeta_t(1-\gamma^*)\widetilde{Z}^*_t=\zeta_t(1-\gamma^*)\vv^\top\vZ^*_t
   \sim  t^{1-\gamma^*}\vv^\top\vZ^*_t = t^{1-\gamma^*}\widetilde{Z}^*_t
\stackrel{a.s./mean}\longrightarrow \widetilde{Z}^{**}_\infty\,,
\end{split}
\end{equation*}
where $\widetilde{Z}_\infty^{**}$ is an integrable strictly positive random variable and $\vZ_t^{**}\stackrel{a.s.}\longrightarrow \widetilde{Z}_\infty^{**}\vu$ 
so that $D_\infty^{**}=\widetilde{Z}^{**}_\infty/\gamma^*$.
In addition, we have
\[\begin{aligned}
t^{-(1-\gamma^*)}E[ \Delta\vM^{**}_{t+1} (\Delta\vM^{**}_{t+1})^\top \mid \mathcal{F}_t]&=
\frac{\zeta_t(1-\gamma^{*})^2}{t^{1-\gamma^*}}E[ \Delta\vM^{*}_{t+1} (\Delta\vM^{*}_{t+1})^\top \mid \mathcal{F}_t]\\
&=
\frac{\zeta_t(1-\gamma^{*})^2}{t^{1-\gamma^*}}
diag(\vZ^{*}_t)(I-diag(\vZ^{*}_t))\\
&=
\frac{\zeta_t(1-\gamma^{*})}{t^{1-\gamma^*}}
diag(\vZ^{**}_t)(I-diag(\vZ^{*}_t))
\stackrel{a.s.}\longrightarrow \Sigma_\infty=\widetilde{Z}^{**}_\infty diag(\vu).
\end{aligned}\]
Therefore, we can  apply 
Theorem~\ref{clt-final}
with $\vA_t=\vZ^{**}_t=\zeta_t(1-\gamma^*)\vZ^*_t$, 
 $\vB_t=\frac{\vD_t^*}{\zeta_t(\gamma^*)}$, $\Phi=\Gamma$, $\phi^{*}=\gamma^{*}$, 
 $\Sigma_\infty=\widetilde{Z}^{**}_\infty diag(\vu)$, 
$\Delta{\vM}_{t+1}=\Delta \vM^{**}_{t+1}=\zeta_{t}(1-\gamma^*)\Delta \vM^*_{t+1}=\zeta_{t}(1-\gamma^*)(\vX^{*}_{t+1}-\vZ^*_t)$.
Finally,  we can conclude by observing that, since $\gamma^*<1$, we have  
 $$
 t^{\gamma^*/2}\left( \frac{\vD_t^*}{t^{\gamma^*}} - \frac{\vD_t^*}{\zeta_t(\gamma^*)} \right) =
 t^{\gamma^*/2} \vD_t^* O\left(1/t^{\gamma^*+1}\right) =
 \frac{\vD^*_t}{t^{\gamma^*}} O\left(1/t^{1-\gamma^*/2}\right)\stackrel{a.s.}\longrightarrow \vzero\,.
 $$
 Note that in 
 Theorem~\ref{clt-final}
 we have the explicit formulas for the elements of the matrix
 $C_{det, \Gamma}$: that is, $C_{det, \Gamma}=1/\gamma^*$ when $N=1$ and
 $$
 C_{det, \Gamma}=\frac{1}{\gamma^*}(\vv^\top diag(\vu)\vv) \vu\vu^\top +
\frac{1}{\widetilde{Z}^{**}_{\infty}}{\mathcal M}^{33}_\infty\,,
$$
where ${\mathcal M}^{33}_\infty$ is the matrix defined in
Theorem~\ref{clt-hat-N}
when $N\geq 2$ 
with the matrix $\Phi$ equal to $\Gamma$. Note that, since the expression
for  ${\mathcal M}^{33}_\infty$ contains the matrix $\Sigma_\infty$, and so
the term $\widetilde{Z}^{**}_{\infty}$, we have that the matrix
$\frac{1}{\widetilde{Z}^{**}_{\infty}}{\mathcal M}^{33}_\infty$ is deterministic.
\end{proof}

\begin{thm}\label{CLT-share-Dstar}  
  Under Assumption \ref{ass-Gamma},  we have
$$
t^{\gamma^*/2}\left(
\frac{\vD_t^*}{\sum_{h=1}^N D^*_{t,h}}-\frac{\vu}{\sum_{h=1}^N u_h}\right)
\stackrel{stably}\longrightarrow 
\mathcal{N}\left(\vzero, 
\frac{1}{D^{**}_\infty}
\frac{\gamma^*}{(\sum_{j=1}^N u_j)^2} 
Q_{\Gamma}C_{det,\Gamma}Q_{\Gamma}^\top\right)\,,
$$
and
$$
\sqrt{\sum_{h=1}^N D^*_{t,h}}
\left(
\frac{\vD_t^*}{\sum_{h=1}^N D^*_{t,h}}-\frac{\vu}{\sum_{h=1}^N u_h}\right)
\stackrel{stably}\longrightarrow 
\mathcal{N}\left(\vzero, 
\frac{\gamma^*}{\sum_{j=1}^N u_j}
Q_{\Gamma}C_{det,\Gamma}Q_{\Gamma}^\top\right)\,,
$$
where $Q_{\Gamma}=I-\vu\vone^\top /\vone^\top\vu$ and
$C_{det,\Gamma}$ is defined in Theorem~\ref{CLT-Dstar}.
Moreover, for each $j$ we have
$$
t^{\gamma^*/2}\left(
\frac{\vD_t^*}{D^*_{t,j}}-\frac{\vu}{u_j}\right)
\stackrel{stably}\longrightarrow 
\mathcal{N}\left(\vzero, 
\frac{1}{D^{**}_\infty}
\frac{\gamma^*}{u_j^2}
Q_{j}C_{det,\Gamma}Q_{j}^\top\right)\,,
$$
and
$$
\sqrt{D^*_{t,j}}
\left(
\frac{\vD_t^*}{D^*_{t,j}}-\frac{\vu}{u_j}\right)
\stackrel{stably}\longrightarrow 
\mathcal{N}\left(\vzero, 
\frac{\gamma^*}{u_j}
Q_{j}C_{det,\Gamma}Q_{j}^\top\right)\,,
$$
where $Q_{j}=I-\vu\ve_j^\top /u_j$ with $\ve_j$ the $j$th vector of the canonical base and
$C_{det,\Gamma}$ is defined in Theorem~\ref{CLT-Dstar}.
\end{thm}
\begin{proof}
  We have to apply
  Corollary~\ref{thm:TCL_ratio}
reported in the appendix
with $\vA_t=\vZ^{**}_t=\zeta_t(1-\gamma^*)\vZ^*_t$,  
 $\vB_t=\frac{\vD_t^*}{\zeta_t(\gamma^*)}$, $\vx=\vone$,
 $\widetilde{A}_\infty=\widetilde{Z}_\infty^{**}=\gamma^* D^{**}_\infty$
 and   $\phi^{*}=\gamma^{*}$. Note that, as specified
 in the proof of this corollary,  we have
 $\widetilde{Z}_\infty^{**}Q_{\Gamma}C_{det,\Gamma}Q_{\Gamma}^\top=
 Q_{\Gamma}{\mathcal M}^{33}_\infty Q_{\Gamma}^\top$, 
where ${\mathcal M}^{33}_\infty$ is the matrix defined in
Theorem~\ref{clt-hat-N}] 
when $N\geq 2$  
with the matrix $\Phi$ equal to $\Gamma$. Similarly, we obtain
the second part of the statement by means of
Corollary~\ref{thm:TCL_ratio} with $\vx=\ve_j$.
\end{proof}
 
\begin{thm}\label{CLT-K}
Under Assumption \ref{ass-W}, we have for a given observed color $c$ in the system
$$
\sqrt{t}\left(\frac{\vK_t(c)}{t} - \widetilde{P}_\infty(c)\vone\right)
\stackrel{stably}\longrightarrow
\mathcal{N}( \vzero,
\widetilde{P}_\infty(c)(1-\widetilde{P}_\infty(c)) C_{det,W} )\,,
$$ 
where $\widetilde{P}_\infty(c)=a.s.-\lim_{t} P_t(h,c)=a.s.-\lim_t K_t(h,c)/t$
for each $h=1,\dots, N$, $\widetilde{P}_\infty(c)\in (0,1)$ a.s.  
and $C_{det,W}$ is a deterministic matrix depending on
the eigen-structure of $W$, 
which is possible to write explicitly.
\end{thm}

\begin{proof} 
For a given color $c$, let $t^*(c)$ be the random time-step of the first extraction of $c$ in the system. Conditionally on $\{t^*(c)<+\infty\}$, for $t\geq t^*(c)$  we can write 
 $$
  P_{t}(h,c)=
  \frac{\sum_{n=1}^t\sum_{j=1}^N w_{j,h}\Delta K_n(j,c)}{\theta_h + t}-
  \frac{\gamma_{j^*(c),h}}{{\theta_h + t}}
 \,,
 $$
 where $\Delta K_{n}(j,c)=K_{n}(j,c)-K_{n-1}(j,c)$ takes values in $\{0,1\}$ and
 $E[\Delta K_{n+1}(j,c)|\mbox{past}]=P_{n}(j,c)$.
Then, we obtain the following dynamics for $P_{t}(h,c)$:
 \begin{equation*}
 P_{t+1}(h,c)=(1-r_{t,h})P_{t}(h,c)+r_{t,h}\sum_{j=1}^N w_{j,h}\Delta K_{t+1}(j,c)\,,
 \end{equation*}
 where $r_{t,h}=1/(\theta_h+t+1)=1/(t+1)+O(1/t^2)$.
  Thus the corresponding vectorial dynamics for 
  $\vP_t(c)=(P_t(1,c),\dots, P_t(N,c))^\top$ is
\begin{equation*}
  \begin{split}
\vP_{t^*(c)}(c)&\neq \vzero,\\
    \vP_{t+1}(c)&=
    \left(1-\frac{1}{t+1}\right)\vP_t(c)+\frac{1}{t+1} W^T {\Delta\vK}_{t+1}(c) + 
    O(1/t^2)\vone\\
    &=
    \vP_t(c) - \frac{1}{t+1}(I-W^\top)\vP_t(c) + \frac{1}{t+1}W^\top \Delta {\vM}_{t+1}(c) +
    O(1/t^2)\vone,
\end{split}
\end{equation*}
where ${\Delta\vK}_t(c)=(\Delta K_t(1,c),\dots, \Delta K_t(N,c))^\top$ and 
$\Delta{\vM}_{t+1}(c)={\Delta\vK}_{t+1}(c)-\vP_t(c)$. 
\\
\indent Considering also the dynamics of  $\vK_t(c)/t$, 
we obtain 
 \begin{equation*}\label{eq-sistema-P}
\begin{split}
\vP_{t+1}(c)&=\vP_t(c) - \frac{1}{t+1}(I-W^\top)\vP_t(c) + \frac{1}{t+1} W^\top \Delta {\vM}_{t+1}(c) + O(1/t^2)\vone\\
\frac{\vK_{t+1}(c)}{t+1}&=\frac{\vK_t(c)}{t} - 
\frac{1}{t+1} \left(\frac{\vK_t(c)}{t}-\vP_t(c)\right) + \frac{1}{t+1} \Delta {\vM}_{t+1}(c)\,. \\   
\end{split}
\end{equation*}
 Moreover, from Theorem~\ref{th-synchro-probab}, we have 
$
\widetilde{P}_t(c)=\vv_W^\top\vP_t(c)\stackrel{a.s.}\longrightarrow \widetilde{P}_\infty(c)\,,
$
and this convergence is  also in mean because  
$(\widetilde{P}_t(c))_t$ is uniformly bounded. 
In addition, we have
\[
E[ \Delta\vM_{t+1} (\Delta\vM_{t+1})^\top \mid \mathcal{F}_t]=
diag(\vP_t)(I-diag(\vP_t))
\stackrel{a.s.}\longrightarrow \Sigma_\infty=\widetilde{P}_\infty(c)(1-\widetilde{P}_\infty(c))I.
\]
  Therefore, we can apply Theorem~\ref{clt-final} with 
 $\vA_t=\vP_t(c)$, $\vB_t=\vK_t(c)/t$, $\Phi=W$, $\phi^{*}=w^*=1$, $\vu=\vu_W=\vone$, $\vv=\vv_W$, 
$\Sigma_\infty=\widetilde{P}_\infty(c)(1-\widetilde{P}_\infty(c))I$, 
  $\Delta{\vM}_{t+1}=\Delta{\vM}_{t+1}(c)={\Delta\vK}_{t+1}(c)-\vP_t(c)$.
  Note that in Theorem~\ref{clt-final} we have the explicit formulas for the elements of the matrix $C_{det,W}$:
  that is, $C_{det, W}=1$ when $N=1$ and
 $$
 C_{det, W}=(\vv_W^\top\vv_W) \vone\vone^\top +
 \frac{1}{\widetilde{P}_\infty(c)(1-\widetilde{P}_\infty(c))}
      {\mathcal M}^{33}_\infty\,,
$$
where ${\mathcal M}^{33}_\infty$ is the matrix defined in
Theorem~\ref{clt-hat-N}] when $N\geq 2$ 
with the matrix $\Phi$ equal to $W$.
Note that, since the expression
for  ${\mathcal M}^{33}_\infty$ contains the matrix $\Sigma_\infty$, and so
the factor $\widetilde{P}_{\infty}(c)(1-\widetilde{P}_\infty(c))$, we have that
the matrix
$\frac{1}{\widetilde{P}_\infty(c)(1-\widetilde{P}_\infty(c))}{\mathcal M}^{33}_\infty$ is deterministic.
\end{proof}

\begin{thm}\label{CLT-ratio-K}
Under Assumption \ref{ass-W}, we have for a given observed color $c$ in the system
\begin{equation*}
\sqrt{t}
\left(\frac{\vK_t(c)}{\sum_{h=1}^N K_t(h,c)}-\frac{1}{N}\vone\right)
\stackrel{stably}\longrightarrow 
         {\mathcal N}\left(\vzero,
         \frac{(1-\widetilde{P}_\infty(c))}{N^2\widetilde{P}_\infty(c)}
         Q_{W} C_{det, W} Q_W^\top\right),
\end{equation*}
and
\begin{equation*}
\sqrt{\frac{\sum_{h=1}^N K_t(h,c)}{1-(Nt)^{-1}\sum_{h=1}^N K_t(h,c)}}
\left(\frac{\vK_t(c)}{\sum_{h=1}^N K_t(h,c)}-\frac{1}{N}\vone\right)
\stackrel{stably}\longrightarrow 
         {\mathcal N}\left(\vzero, \frac{1}{N}
         Q_WC_{det, W} Q_W^\top\right).
\end{equation*} 
where $ Q_W=(I-N^{-1}\vone\vone^\top)$ and $C_{det,W}$ is defined in
Theorem~\ref{CLT-K}.
\end{thm}
\begin{proof}
  We  have to apply Corollary~\ref{thm:TCL_ratio}
reported in the appendix
with $\vA_t=\vP_t(c)$, $\vB_t=\vK_t(c)/t$,
  $\vx=\vone$, $\widetilde{A}_\infty=\widetilde{P}_\infty(c)$,
  $\phi^{*}=w^{*}=1$ and $\vu=\vu_W=\vone$.
Note that, as specified
in the proof of this corollary,  we have
$\widetilde{P}_\infty(c)(1-\widetilde{P}_\infty(c))Q_{W}C_{det,W}Q_{W}^\top=
 Q_{W}{\mathcal M}^{33}_\infty Q_{W}^\top$, 
where ${\mathcal M}^{33}_\infty$ is the matrix defined in
Theorem~\ref{clt-hat-N}  
with the matrix $\Phi$ equal to $W$.
\end{proof}

\begin{rem}\label{rem:CLT when assumption second eigenvalue}
 (Discussion of the Assumption~\ref{ass-Gamma} and Assumption~\ref{ass-W})\\ 
As observed in Remark~\ref{suppl:rem:diag-irrid} in the appendix, removing the diagonalizability of the matrices $\Gamma$ and $W$ leads to analogous results (cf.~\cite{yang2025}). 
Moreover, in order to deal with reducible matrices, we  refer 
to~\cite{ale-cri-ghi-innovation-2023, ale-cri-ghi-complete, ale-ghi}.
Finally, when ${\mathcal Re}(\gamma^*_2)/\gamma^*\geq 1/2$, as highlighted in  
Remark~\ref{rem:divergence assumption second eigenvalue} in appendix, 
the rate of convergence is slower than $t^{\gamma^*/2}$. Indeed,
similarly to the stochastic approximation literature (e.g. \cite{Zhang2016}), 
for the case 
 ${\mathcal Re}(\gamma^*_2)/\gamma^*=1/2$ we expect a CLT with a logarithmic correction in the scaling; while 
 in the case ${\mathcal Re}(\gamma^*_2)/\gamma^*>1/2$, a power-law rate with an exponent smaller than $\gamma^*/2$ could hold true, but not a Gaussian limit. 
 Analogous considerations can be done for the cases ${\mathcal Re}(w^*_2)=1/2$ and ${\mathcal Re}(w^*_2)>1/2$.
\end{rem}

\section{Statistical inference}\label{stat-inference}

  We here present some statistical tools based on the results of
  Section~\ref{main-CLTs}. More precisely, we provide
  a statistical test for each of the two interaction matrices,
$\Gamma$ and $W$, based on the observable processes
$\vD^*_t$ and $\vK_t(c)$. Furthermore, we use the latter
process to provide also a confidence
interval  for the limit probability $\widetilde{P}_\infty(c)$.
The general procedure to be followed in order to obtain these statistical instruments
is described in
Section~\ref{app:statistical_tools} and
Section~\ref{app:statistical_tools_examples}. 
In particular, in our framework, we have to apply it 
with $\vB_t=\frac{\vD_t^*}{\zeta_t(\gamma^*)}$
(or, equivalently $\vB_t=\frac{\vD_t^*}{t^{\gamma^*}}$) and with 
$\vB_t=\vK_t(c)/t$. To this end, we observe that  
Assumption~\ref{ass-matrix-Sigma} is true in both cases with
$\Sigma_{det}=\Sigma_{det,\Gamma}=diag(\vu)$ in the first one and
$\Sigma_{det}=\Sigma_{det,W}=I$ in the second one.
\\

\indent We firstly focus on the case $N=2$ and, then, we consider
a particular type of interaction (mean-field interaction) in the general case
$N\geq 2$. It is worthwhile to underline that the statistical tools
described in Section~\ref{app:statistical_tools}
are very general and so it is also possible to
 explicit the formulas for any family of interaction matrices satisfying the required assumptions.

\subsection{The case $N = 2$}\label{sec:case-N2}

\subsubsection{Hypothesis test on $\Gamma$}
In what follows we assume the parameters $\gamma^*\in (0,1)$ and $r=u_1/u_2>0$ to be known. 
\\

We use the re-parametrization of the matrix $\Gamma$
given by the constrains
$$
r\gamma_{1,1}+\gamma_{2,1}=r\gamma^*
\quad\mbox{and}\quad
r\gamma_{1,2}+\gamma_{2,2}=\gamma^*\,.
$$
and make inference on the other two parameters $\gamma_{1,2}$ and $\gamma_{2,1}$
(that have to be such that the assumptions on the matrix $\Gamma$ are verified).
If we use the parametrization $\iota=\gamma_{1,2}+\gamma_{2,1}$
and $\eta=\tfrac{\gamma_{1,2}}{\gamma_{1,2}+\gamma_{2,1}}$
so that $\iota$ rules the intensity of the whole interaction between the
two processes  due to matrix $\Gamma$,
while $\eta$ (resp. $(1-\eta)$) denotes the percentage of this intensity
due to the influence of process $1$ on process $2$  
(resp., of process $2$ on process $1$), the expressions for
the elements of the matrix $\Gamma$ become
\begin{equation}\label{eq:elements_Gamma_N_equal_2}
\gamma_{1,1}=\gamma^*-\frac{(1-\eta)\iota}{r},
\quad
\gamma_{1,2}=\eta\iota,
\quad
\gamma_{2,1}=(1-\eta)\iota
\quad\mbox{and}\quad
\gamma_{2,2}=\gamma^*-r\eta\iota,
\end{equation}
where (by the required conditions on $\Gamma$)  
\begin{equation}\label{eq:conditions-par}
\begin{aligned}
 &\eta\in (0,1)\quad\mbox{and}\quad
 \iota \in J_\eta\quad\mbox{with},\\
 & J_\eta = \left( 0,
 \min\big(
 \tfrac{\gamma^*r}{1-\eta}, \tfrac{\gamma^*}{r\eta}\big) \right]
\cap
\left( 0,
 \tfrac{1-\gamma^*}{\eta(1-r)}I_{(r<1)}+
 \tfrac{1-\gamma^*}{(1-\eta)(1-1/r)}I_{(r>1)}
 +\infty I_{(r=1)} \right).
 \end{aligned}
 \end{equation}
 \paragraph*{\bf Two-sided test} If we take the null hypothesis
$H_0:\;\iota=\iota_0\,,\,\eta=\eta_0$,  
where $\iota_0$ and $\eta_0$ satisfy
\eqref{eq:conditions-par}, i.e. $\iota_0\in J_{\eta_0}$, and
the required condition in
  Assumption~\ref{ass-Gamma} for the
second eigenvalue of $\Gamma$, that is 
$\iota_0\frac{(1-\eta_0)+r^2\eta_0}{r}>\gamma^*/2$, then
we can use Theorem~\ref{CLT-Dstar} and obtain 
  the test statistic (see
  \eqref{test} from Remark~\ref{rem:N_equal_2_Sigma_diagonal} 
  with $\phi=\gamma^*$,
$\phi_2^*=\gamma^*_2=
  \gamma^*-[(1-\eta_0)\iota_0+r^2\eta_0\iota_0]/r$, $g(x)=x$, 
 and
$\vB_t=\frac{\vD_t^*}{\zeta_t(\gamma^*)}$
 or, equivalently $\vB_t=\frac{\vD_t^*}{t^{\gamma^*}}$)
\begin{equation}\label{eq:test_statistics_Gamma}
q_{r,\eta_0}\Delta_0
\,\frac{(D^*_{1,t}-rD^*_{2,t})^2}{\widetilde{D}^*_t}
\stackrel{d}\longrightarrow \chi^2(1)
\quad\mbox{under } H_0\,,
\end{equation}
where
$\widetilde{D}^*_t=\vv^\top \vD^*_t=\frac{\eta_0 r D^*_{1,t}+(1-\eta_0)D^*_{2,t}}{\eta_0 r +(1-\eta_0)}$,
$q_{r,\eta_0} = \frac{2[(1-\eta_0)+\eta_0 r^2]}{r(1+r)[(1-\eta_0)+\eta_0r]}$ 
and 
$$
\Delta_0 =\frac{1}{2}-\frac{\gamma^{*}_2}{\gamma^{*}}= 
\frac{\iota_0}{\gamma^{*}}\frac{(1-\eta_0)+\eta_0r^2}{r}-\frac{1}{2}\,.
$$
It is immediate to see that the test statistic \eqref{eq:test_statistics_Gamma}
has an increasing dependence on the parameter $\iota_0$, while
its behaviour with respect the parameter $\eta_0$ is not clear.
However, we can observe that it can be rewritten as
$$
  \frac{2}{r(1+r)}
  \frac{(1-\eta_0)+\eta_0 r^2}{(1-\eta_0)+\eta_0 r D^*_{1,t}/D^*_{2,t}}
  \Delta_0
   D^*_{2,t}(D^*_{1,t}/D^*_{2,t}-r)^2\,,
$$
where, for any value of $\eta_0$, we have 
  $$
  \frac{(1-\eta_0)+\eta_0 r^2}{(1-\eta_0)+\eta_0 r D^*_{1,t}/D^*_{2,t}}
  \stackrel{a.s.}\longrightarrow 1\,.
  $$
Hence, we can simplify \eqref{eq:test_statistics_Gamma} as
\begin{equation}\label{eq:test_statistics_Gamma-2}
\tfrac{2}{r(1+r)} \Delta_0 D^*_{2,t}(D^*_{1,t}/D^*_{2,t}-r)^2
\stackrel{d}\longrightarrow \chi^2(1)
\quad\mbox{under } H_0\,, 
\end{equation}
where the dependence on $\iota_0$ and $\eta_0$ is only in the
factor $\Delta_0$.   
Notice also that in the present framework the two eigenvalues 
$\gamma^{*}$ and $\gamma_2^{*}$ are both real and so 
the quantity $\Delta_0$ represents the distance of the ratio between
the second and the leading eigenvalue from the critical value $1/2$ 
(see condition ${\mathcal Re}(\gamma^*_2)/\gamma^*<1/2$ in Assumption~\ref{ass-Gamma}). Moreover, $\Delta_0$ is the only quantity in
the test statistics~\eqref{eq:test_statistics_Gamma-2} that
contains the parameters $\iota_0$ and $\eta_0$.  
Moreover, it is  neither random nor time-dependent.
 The random quantity
$D^*_{2,t}(D^*_{1,t}/D^*_{2,t}-r)^2=[\sqrt{D^*_{2,t}}(D^*_{1,t}/D^*_{2,t}-r)]^2$
 describes the fluctuations of $D^*_{1,t}/D^*_{2,t}$ around $r$
   (note that \eqref{eq:test_statistics_Gamma-2} is coherent with the
   convergence result stated in~Theorem~\ref{CLT-share-Dstar}).
   
\paragraph*{\bf One-sided test} Since $\Delta_0$,
 and so the statistic~\eqref{eq:test_statistics_Gamma-2}, 
  is increasing in $\iota_0$, we can replace 
  the condition $\iota=\iota_0$ in $H_0$  
  by $\iota\geq\iota_0$. Further, the quantity $\Delta_0$
  is increasing in $\eta_0$ when $r>1$ and decreasing in $\eta_0$ when $r<1$.
  Hence, the statistic~\eqref{eq:test_statistics_Gamma-2} works well also
  if we replace the condition $\eta=\eta_0$ in $H_0$ by
  $\eta \geq \eta_0$ when $r>1$ or by $\eta \leq \eta_0$ when $r<1$.
  Therefore, if we set $\eta_0=1/2$, we get 
$$
\Delta_0 =
\frac{1}{2}-
\frac{\gamma^*_2}{\gamma^*}=
\frac{\iota_0}{\gamma^*}\frac{1+r^2}{2r}-\frac{1}{2}
$$
and we can perform a one-sided test with
  one of the following two null hypotheses, according to the value of~$r$: 
\begin{itemize}
\item Case $r>1$:
  $H_0:\;\iota\geq \iota_0,\, \eta\geq 1/2$
  (the interaction intensity is at least $\iota_0$ and
  the processes equally influence each other or 
  the interaction is mostly due to the influence of process $1$ 
  on process $2$), where  
  $\iota_0$ satisfies \eqref{eq:conditions-par} for some $\eta$ in $H_0$,
  i.e. $\iota_0\in \cup_{\eta\geq 1/2} J_{\eta}$, and
$\iota_0>\gamma^*\frac{r}{1+r^2}$;\\
\item Case $r<1$: $H_0:\;\iota\geq \iota_0,\, \eta\leq 1/2$
  (the interaction intensity is at least $\iota_0$ and
    the processes equally influence each other or 
  the interaction is mostly due to the influence of process $2$ 
  on process $1$), where 
$\iota_0$ satisfies \eqref{eq:conditions-par} for some $\eta$ in $H_0$, i.e. $\iota_0\in \cup_{\eta\leq 1/2} J_{\eta}$, and
$\iota_0>\gamma^*\frac{r}{1+r^2}$.
\end{itemize}
Finally, in the very special case when $r=1$,
the test statistic~\eqref{eq:test_statistics_Gamma-2} 
does not depend on the value of $\eta_0$ and it can be used for
testing the intensity $\iota$ only. 

\paragraph*{\bf Power} Following 
Remark~\ref{rem-power}, 
  we find that, under  the alternative 
  $H_1:\,\iota=\iota_1,\, \eta=\eta_1$, with 
  $\iota_1\in J_{\eta_1}$ 
   and 
$\iota_1\frac{(1-\eta_1)+r^2\eta_1}{r}>\gamma^*/2$, 
the statistic~\eqref{eq:test_statistics_Gamma-2}
is asymptotically distributed as 
$$
 \frac{\Delta_0}{\Delta_1}\,
 \chi^2(1)=
 \frac{\frac{\iota_0}{\gamma^*}\frac{(1-\eta_0)+\eta_0 r^2}{r}-\frac{1}{2}}
      {\frac{\iota_1}{\gamma^*}\frac{(1-\eta_1)+\eta_1 r^2}{r}-\frac{1}{2}}\,
 \chi^2(1)\,.
$$
Notice that
 the term $\tfrac{\Delta_0}{\Delta_1}$ is 
 time-invariant and this means that increasing the time $t$ does not
 increase the power.
 Nevertheless, the power
 increases as $\eta_1$ decreases if $r>1$
 or as $\eta_1$ increases if $r<1$.
 In addition, the power improves as the value of $\iota_1$ decreases
 and it tends toward $1$ as $\Delta_1\to 0$, that is as $\iota_1$
 approaches its minimum admissible value
 $\tfrac{r}{(1-\eta_1)+\eta_1 r^2}\frac{\gamma^*}{2}$. 
 Moreover, when $\iota_1$
 is even lower or equal to
 $\tfrac{r}{(1-\eta_1)+\eta_1 r^2}\frac{\gamma^*}{2}$, i.e. when condition ${\mathcal Re}(\gamma^*_2)/\gamma^*< 1/2$ does not hold,
 the power is asymptotically equal to 1 because the test statistic~\eqref{eq:test_statistics_Gamma-2} diverges to infinity as a consequence of Remark~\ref{rem:CLT when assumption second eigenvalue}.

\subsubsection{Hypothesis test on $W$}
As $W^\top\vone=\vone$ (i.e. $w^*=1$ and $r=1$),
we have 
\begin{equation*}
w_{1,1}=1-w_{2,1}
\quad\mbox{and}\quad
w_{2,2}=1-w_{1,2}
\end{equation*}
and so we can make inference on the two parameters
$w_{2,1}$ and $w_{1,2}$.
  Using the  parametrization $\iota=w_{1,2}+w_{2,1}$ for the intensity of
  the interaction due to the matrix $W$,
and $\eta=\tfrac{w_{1,2}}{w_{1,2}+w_{2,1}}$ (resp. $(1-\eta)$) for
the percentage of this intensity due to the influence of process $1$ 
on process $2$ 
(resp., of process $2$ on process $1$), we have 
\begin{equation}\label{eq:elements_W_N_equal_2}
w_{1,1}=1-(1-\eta)\iota,
\qquad
w_{1,2}=\eta\iota,
\qquad
w_{2,1}=(1-\eta)\iota
\qquad\mbox{and}\qquad
w_{2,2}=1-\eta\iota\,,
\end{equation}
where (by the required conditions on $W$) 
\begin{equation}\label{eq:conditions-parameters-W}
  \eta\in (0,1)\qquad\mbox{and}\qquad
  \iota\in\big(0, \min(1/\eta, 1/(1-\eta)\big)
  \subseteq (0,2)\,.
\end{equation}
\paragraph*{\bf Two-sided test} If we take the null hypothesis
$H_0:\;\iota=\iota_0\,,\,\eta=\eta_0$, 
where \eqref{eq:conditions-parameters-W} and
$\iota_0>1/2$ (the required condition in Assumption~\ref{ass-W} for the 
second eigenvalue of $W$) are satisfied, then
we can use Theorem~\ref{CLT-K} and obtain 
the test statistic (see \eqref{test}
from Remark~\ref{rem:N_equal_2_Sigma_diagonal} 
with $\phi=w^*=1$, $r=1$, $\phi_2^*=1-\iota_0$,  $g(x)=x(1-x)$ 
and
$\vB_t=\vK_t(c)/t$ with $c$ an arbitrary old color)
\begin{equation}\label{eq:test_statistics_W}
\begin{aligned}
&\Delta_0
\,\frac{(K_{1,t}(c)-K_{2,t}(c))^2}{\widetilde{K}_t(c)(1-\widetilde{K}_t(c)/t)}
\stackrel{d}\longrightarrow \chi^2(1)
\quad\mbox{under } H_0\,,\\
&\mbox{where}\qquad\Delta_0 =\frac{1}{2}-(1-\iota_0)=\iota_0-\tfrac{1}{2},
\end{aligned}
\end{equation}
and
\begin{equation}\label{def:K_tilde}
\widetilde{K}_t(c)=\vv^\top \vK_t(c)=\eta_0 K_{1,t}(c)+(1-\eta_0)K_{2,t}(c).
\end{equation}
Since $K_{j,t}(c)/t\stackrel{a.s.}\to \widetilde{P}_\infty(c)$, we 
obtain for any $\eta_0$ and $\eta_1$ 
$$
\frac{\eta_1 K_{1,t}(c)/t+(1-\eta_1)K_{2,t}(c)/t}{\widetilde{K}_t(c)/t}\,
\frac{(1- \eta_1 K_{1,t}(c)/t+(1-\eta_1)K_{2,t}(c)/t)}{(1-\widetilde{K}_t(c)/t)}
\stackrel{a.s.}\longrightarrow 1\,,
$$
and so the choice of the value of $\eta_0$ in $H_0$ does not affect the asymptotic distribution of the test statistic. This means that 
  we can use the test statistic~\eqref{eq:test_statistics_W} with an arbitrary
  value of $\eta_0$ in order to perform a test on the intensity $\iota$.
In particular, if we choose $\eta_0=1/2$, we simply have
$\widetilde{K}_t(c)=\tfrac{K_{1,t}(c)+K_{2,t}(c)}{2}$.

\paragraph*{\bf One-sided test} The statistic~\eqref{eq:test_statistics_W} is 
  increasing in $\iota_0$ and so it works well also
  for a one-sided test with the 
    condition $\iota=\iota_0$ in $H_0$ 
  replaced by $\iota\geq \iota_0$, where $\iota_0\in (\tfrac{1}{2},2)$.   

  \paragraph*{\bf Power} Following 
  Remark~\ref{rem-power}, 
  we find that, under the alternative 
  $H_1:\,\iota=\iota_1$, where $\iota_1\in(\tfrac{1}{2},2)$ with 
  $\iota_1\neq \iota_0$, 
 the statistic~\eqref{eq:test_statistics_W} is asymptotically distributed as 
$$
 \frac{\Delta_0}{\Delta_1} \chi^2(1)=
 \frac{\iota_0-\tfrac{1}{2}}{\iota_1-\tfrac{1}{2}}\,
 \chi^2(1)\,.
$$
Notice that the term $\tfrac{\Delta_0}{\Delta_1}$ is time-invariant and so,
 as before, increasing the time $t$ does not increase the power.
 However, the power is monotone decreasing in $\iota_1$, and it tends to $1$
 as $\iota_1$ approaches its lower bound $1/2$, and it is equal to one for $\iota_1\leq 1/2$ by Remark~\ref{rem:CLT when assumption second eigenvalue}.

  \subsubsection{Confidence interval for $\widetilde{P}_\infty(c)$}
\begin{figure}
    \begin{center}
   \includegraphics[scale=0.25]{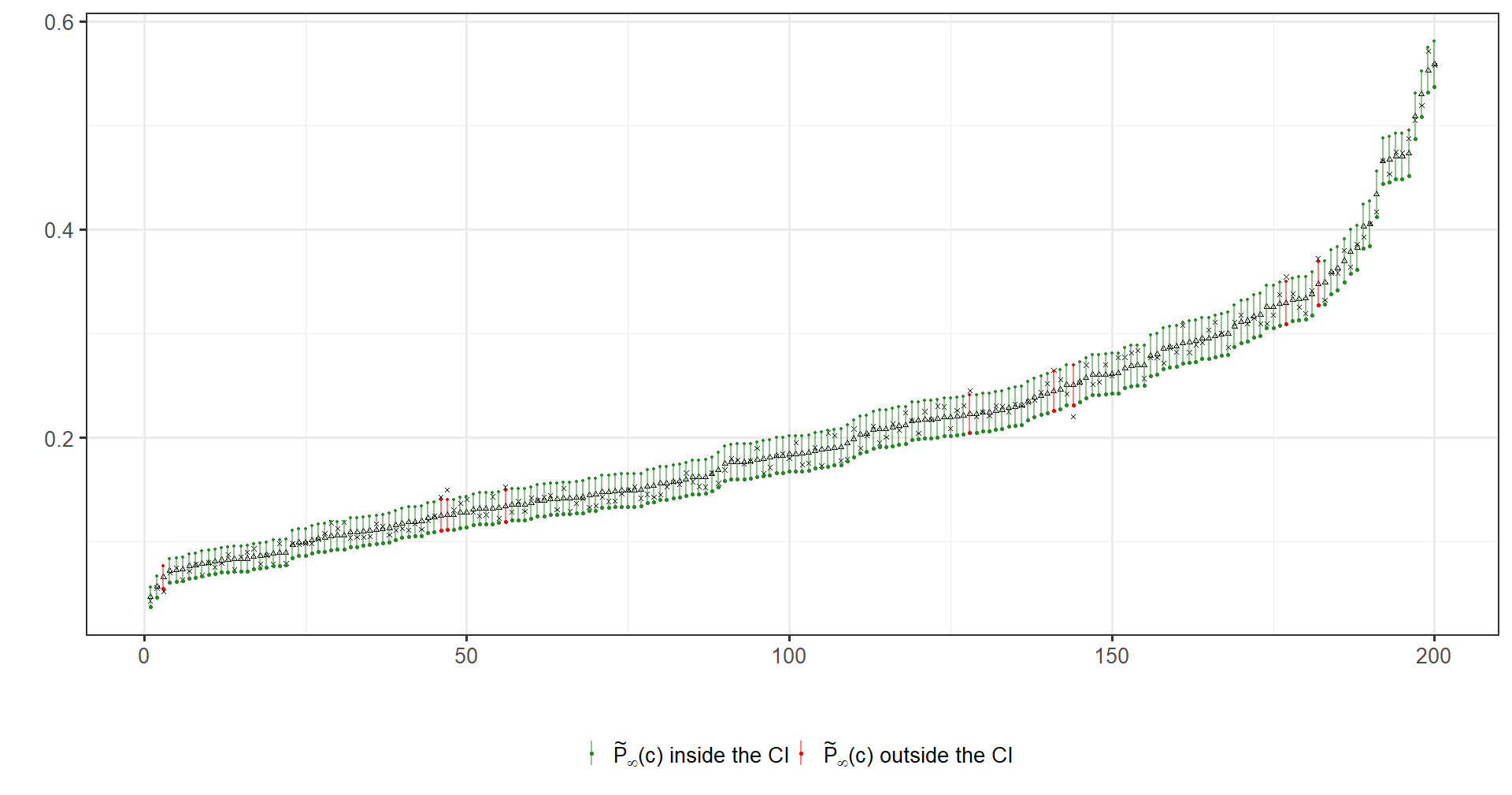} 
   \caption{Confidence interval (see~\eqref{eq:confidence_interval_N=2}) 
     for the random variable 
     $\widetilde{P}_\infty(c)$
     at time-step $t=10^3$ and related to item $c$ with highest
     $\vone^\top\vK_t(c)$  in the case $N=2$,
     $\Gamma$ defined as in~\eqref{eq:elements_Gamma_N_equal_2}
     with $r=0.75$, $\gamma^{*}=0.75$, $\eta=1/2$ and $\iota_{\Gamma}=1$, and
     $W$ defined as
     in~\eqref{eq:elements_W_N_equal_2} with $\eta=1/2$ and $\iota_W=1.25$.
     Number of simulations $S=200$
     (sorted on the $x$-axis according to the value of $\vone^\top\vK_t(c)$).
     The crosses represent the value (reported on the $y$-axis) of
     $\widetilde{P}_\infty(c)$ 
     estimated by
     $\vone^\top\vK_{t_\infty}(c)/(Nt_\infty)$ at time-step $t_{\infty}=10^5$.
     }
   \label{Figure:confidence_intervals_symmetric}
  \end{center}
\end{figure}

Fix the confidence level $(1-\alpha)$ and denote by $z_\alpha$ the quantile of
order $(1-\alpha/2)$ of the standard normal distribution, that is 
${\mathcal N}(0,1)(z_\alpha,+\infty)=\alpha/2$.
Assuming $\iota>1/2$,
we can use Theorem~\ref{CLT-K} in order to construct a confidence interval
for the random limit $\widetilde{P}_\infty(c)$
associated to a given item $c$ 
(see the interval \eqref{conf-interval}
from Remark~\ref{rem:N_equal_2_Sigma_diagonal}   
with $\phi=w^*=1$, $r=1$, $g(x)=x(1-x)$ 
and 
$\vB_t=\vK_t(c)/t$). Precisely, we obtain:
\begin{equation}\label{eq:confidence_interval_N=2}
CI_{1-\alpha}(\widetilde{P}_\infty(c))\ =\
\frac{\widetilde{K}_t(c)}{t} \pm\ \frac{z_{\alpha}}{t^{1/2}}
\!\sqrt{ \frac{\widetilde{K}_t(c)}{t}\left(1-\frac{\widetilde{K}_t(c)}{t}\right)c_{\eta}},
\end{equation}
with $c_{\eta} = \eta^2+(1-\eta)^2$ and $\widetilde{K}_t(c)$ defined as in~\eqref{def:K_tilde}. It is interesting to notice the absence of the dependence on the specific value of $\iota$.
On the other hand, while the dependence of
  $\widetilde{K}_t(c)$ on $\eta$ is not important, because 
  $K_{j,t}(c)\stackrel{a.s.}\to \widetilde{P}_\infty(c)$ so that
  $\frac{\eta_1 K_{1,t}(c)/t+(1-\eta_1)K_{2,t}(c)/t}{\widetilde{K}_t(c)/t}
  \stackrel{a.s.}\to 1$, the presence of the quantity $c_{\eta}$
  implies that
  we need to know the value of the parameter $\eta$. However, $c_{\eta}$ is
  always smaller than $1$ for $\eta\in (0,1)$ and hence, if we do not know
  the value of $\eta$, we can use $\widetilde{K}_t(c)$ with
  $\eta=1/2$, so that we have
  $\widetilde{K}_t(c)=\tfrac{K_{1,t}(c)+K_{2,t}(c)}{2}$, and
  take the largest interval obtained
  from~\eqref{eq:confidence_interval_N=2} with $c_{\eta}$ replaced
  by $1$.
\\As an example,
we have
simulated $S=200$ independent pair of innovation processes with $\Gamma$ and $W$
defined with $r=0.75$, $\gamma^{*}=0.75$, $\eta=1/2$, $\iota_{\Gamma}=1$ and
$\iota_{W}=1.25$ and then,
for any simulation, we have constructed 
the confidence intervals~\eqref{eq:confidence_interval_N=2} of
level $(1-\alpha)=95\%$ 
computed at time-step $t=10^3$ associated to the item $c$
with highest $(K_{t,1}(c)+K_{t,2}(c))$
(recall that the interval has the same expression
for any given item $c$).
In Figure~\ref{Figure:confidence_intervals_symmetric} we can see these $S=200$
confidence intervals (sorted according to the value of  
$K_{t,1}(c)+K_{t,2}(c)\,$) and the estimates of the corresponding
$\widetilde{P}_\infty(c)$ computed as
$\tfrac{K_{t_{\infty},1}(c)+K_{t_{\infty},2}(c)}{2t_\infty}$ at time-step
$t_{\infty}=10^5$.
The empirical coverage is represented by the proportion of times the
estimates of $\widetilde{P}_\infty(c)$ belong to the corresponding
confidence interval, and in this simulation this is equal to $0.955$, 
which is very close to the nominal value of $0.95$.

\subsection{Mean-field interaction for $N\geq 2$}\label{sec:mean_field_case}

\subsubsection{Hypothesis test on $\Gamma$}
Suppose that $\Gamma$  is of the form
\begin{equation}\label{eq:elements_Gamma_mean-field}
\gamma_{j,h}=\phi\left(\iota/N+\delta_{j,h}(1-\iota)\right)
\qquad \phi,\iota\in (0,1]\,.
 \end{equation}
We refer to this kind of interaction matrix as the mean-field interaction 
 because in this case the term $\sum_{j=1}^N \gamma_{j,h}D^*_{t,j}$ in \eqref{birth-prob-inter} 
 becomes 
 $\phi\left(\iota \frac{1}{N}\sum_{j=1}^N D^*_{t,j} + (1-\iota) D^*_{t,h}\right)$
and so the conditional probability $Z^*_{t,h}$ depends on a convex combination between $D^*_{t,h}$ and 
the averaged value of the all $D^*_{t,j}$ in the system. Note that, for this matrix $\Gamma$, the leading eigenvalue $\gamma^*$ coincides with the parameter $\phi$.
\\

We can make
  inference on the parameter $\iota$ that rules the intensity
  of the interaction.
  
\paragraph*{\bf Two-sided test} If we take the null hypothesis 
 $H_0:\;\iota=\iota_0$, where $\iota_0\leq 1$ and $\iota_0 >1/2$  
(the required condition in Assumption~\ref{ass-Gamma} for the 
  second eigenvalue of $\Gamma$), then 
  we can use Theorem~\ref{CLT-Dstar} and obtain 
  the test statistic (see \eqref{test-mean} from Remark~\ref{rem:mean_field_Sigma_diagonal}  
with $\phi=\gamma^*$, $\phi_2^{*}=\gamma^*(1-\iota_0)$, $g(x)=x$, 
and
$\vB_t=\frac{\vD_t^*}{\zeta_t(\gamma^*)}$
 or, equivalently $\vB_t=\frac{\vD_t^*}{t^{\gamma^*}}$)
\begin{equation}\label{eq:test_statistics_mean_field_case_Gamma}
\begin{aligned}
&2\Delta_0\frac{\big\|\vD^*_t-\widetilde{D}^*_t\vone
  \big\|^2}{\widetilde{D}^*_t}  
\stackrel{d}\longrightarrow\chi^2(N-1)\quad\mbox{under } H_0,\\
&\mbox{where}\qquad
\Delta_0=\frac{1}{2}-\frac{\gamma^{*}(1-\iota_0)}{\gamma^{*}}=\iota_0-\frac{1}{2}
\end{aligned}
\end{equation}
and $\widetilde{D}^*_t=\vv^\top \vD^*_t=\tfrac{\vone^\top\vD^*_t}{N}$.
Note that this test statistic does not depend on the parameter $\phi=\gamma^*$ and so the value of this parameter is not needed for the test on $\iota$.

\paragraph*{\bf One-sided test} Since the test statistic~\eqref{eq:test_statistics_mean_field_case_Gamma} is  increasing in $\iota_0$,  it works well also
for the one-sided test with $H_0:\;\iota\geq \iota_0$, 
  where $\iota_0\in (1/2,1]$.

\paragraph*{\bf Power}  Following 
Remark~\ref{rem-power},  
we observe that, under the alternative $H_1:\, \iota=\iota_1$, 
with $\iota_1\in (1/2,1]$,  
the statistics~\eqref{eq:test_statistics_mean_field_case_Gamma} is
asymptotically distributed as
$$\frac{\Delta_0}{\Delta_1}\chi^2(N-1)=
\frac{\iota_0 -\tfrac{1}{2}}{\iota_1-\tfrac{1}{2}} \chi^2(N-1)\,.
$$
  Therefore, as before, increasing the time $t$ does not increase the power.
Moreover, the power is monotone decreasing in $\iota_1$,
  and it tends to $1$ as $\iota_1$ approaches its lower bound $1/2$, being equal to 1 for $\iota_1\leq 1/2$ by Remark~\ref{rem:CLT when assumption second eigenvalue}.

\subsubsection{Hypothesis test on $W$}
Suppose that $W$ is of the form
\begin{equation}\label{eq:elements_W_mean-field}
w_{j,h}=\iota/N+\delta_{j,h}(1-\iota)
\qquad \iota \in (0,1]\,.
\end{equation}
\paragraph*{\bf Two-sided test} If we take the null hypothesis  
$H_0:\;\iota=\iota_0$,  where $\iota_0\leq 1$ and $\iota_0>1/2$
(the required condition in Assumption~\ref{ass-W} for the 
second eigenvalue of $W$), then we can use Theorem~\ref{CLT-K} and obtain 
the test statistic (see \eqref{test-mean} from Remark~\ref{rem:mean_field_Sigma_diagonal}  
with $\phi=w^*=1$, $\phi_2^{*}=w^*(1-\iota_0)=(1-\iota_0)$, $g(x)=x(1-x)$, 
and
$\vB_t=\vK_t(c)/t$)
\begin{equation}\label{eq:test_statistics_mean_field_case_W}
\begin{aligned}
&2\Delta_0
\frac{\big\|\vK_t(c)-\widetilde{K}_t(c)\vone
  \big\|^2}{\widetilde{K}_t(c)(1-\widetilde{K}_t(c)/t)}  
\stackrel{d}\longrightarrow \chi^2(N-1)\quad\mbox{under } H_0,\\
&\mbox{where}\qquad
\Delta_0=\frac{1}{2}-(1-\iota_0)=\iota_0-\frac{1}{2},
\end{aligned}
\end{equation}
$c$ is an arbitrary observed item and $\widetilde{K}_t(c)=\vv^\top \vK_t(c)=\tfrac{\vone^\top\vK_t(c)}{N}$.

\paragraph*{\bf One-sided test}
The test statistic~\eqref{eq:test_statistics_mean_field_case_W} is increasing in $\iota_0$ and so it works well also
for the one-sided test with
$H_0: \;\iota\geq \iota_0$, where $\iota_0\in (1/2,1]$.

\paragraph*{\bf Power} Following 
Remark~\ref{rem-power},  
we observe that, under the alternative $H_1:\, \iota=\iota_1$, with
$\iota_1\in (1/2,1]$,
  the statistics~\eqref{eq:test_statistics_mean_field_case_W}
is asymptotically distributed exactly as the
previous statistics~\eqref{eq:test_statistics_mean_field_case_Gamma}
and so the same considerations on the power of the test hold true.

\subsubsection{Confidence interval for $\widetilde{P}_\infty(c)$}
Fix the confidence level $(1-\alpha)$ and denote by $z_\alpha$ the quantile of
order $(1-\alpha/2)$ of the standard normal distribution, that is 
${\mathcal N}(0,1)(z_\alpha,+\infty)=\alpha/2$. Assuming $\iota\in (1/2,1]$,
we can use Theorem~\ref{CLT-K} in order to
  construct a confidence interval
for the random limit $\widetilde{P}_\infty(c)$
associated to a given item $c$ 
(see the interval \eqref{conf-interval-mean} from Remark~\ref{rem:mean_field_Sigma_diagonal} 
with $\phi=w^*=1$, $g(x)=x(1-x)$ 
and
$\vB_t=\vK_t(c)/t$). Precisely, we obtain:
\begin{equation}\label{eq:confidence_interval_mean_field_case}
CI_{1-\alpha}(\widetilde{P}_\infty(c))\ =\
\frac{\widetilde{K}_t(c)}{t} \pm\ \frac{z_{\alpha}}{t^{1/2}}
\!\sqrt{ \frac{\widetilde{K}_t(c)}{t}\left(1-\frac{\widetilde{K}_t(c)}{t}\right)\frac{1}{N}},
\end{equation}
where $\widetilde{K}_t(c)=\vv^\top \vK_t(c)=\tfrac{\vone^\top\vK_t(c)}{N}$.
Note that this interval does not depend on the value of
the parameter $\iota$.\\

\begin{figure}
    \begin{center}
   \includegraphics[scale=0.25]{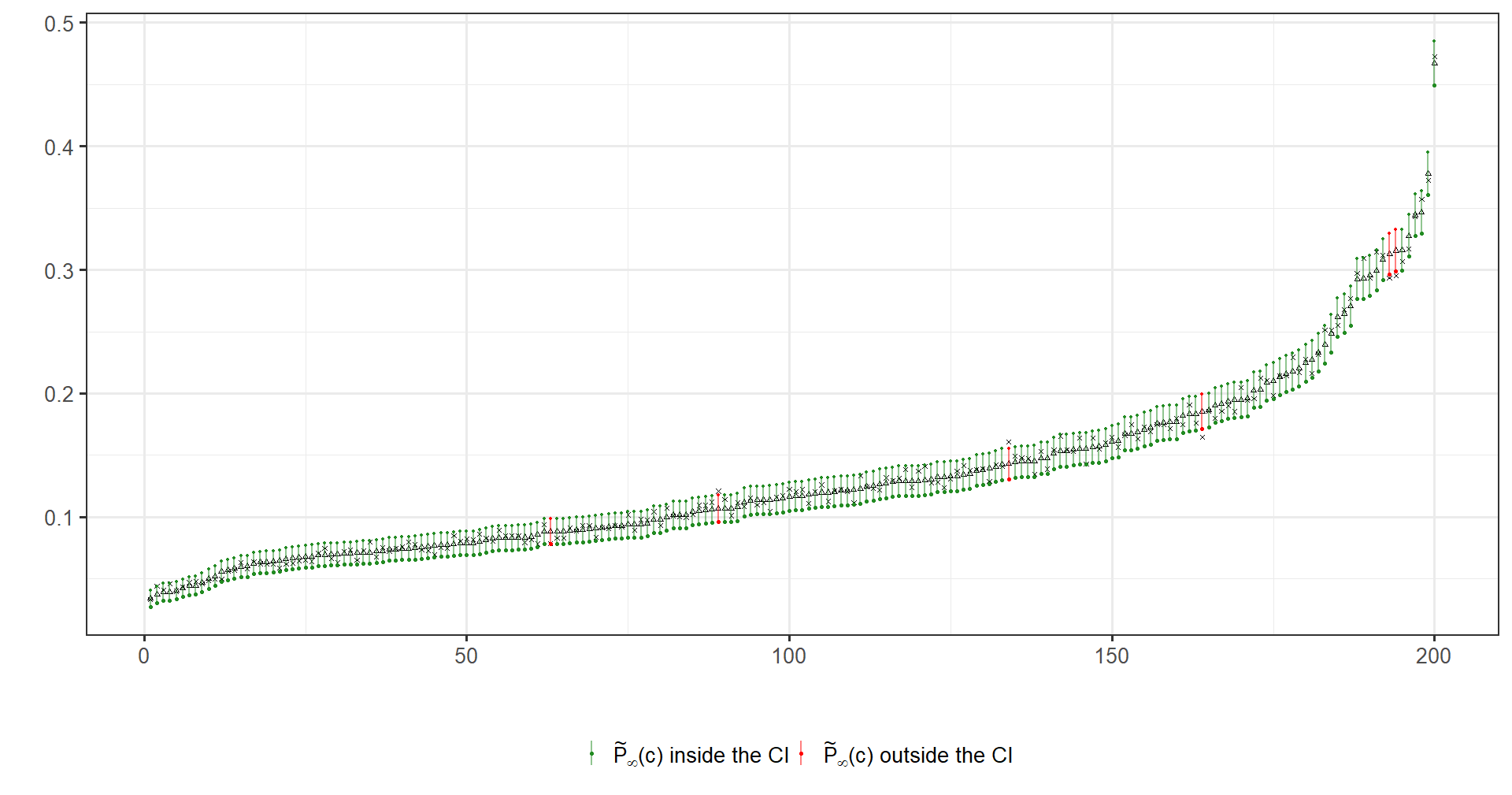} 
   \caption{Confidence interval
     (see~\eqref{eq:confidence_interval_mean_field_case})
     for the random variable
     $\widetilde{P}_\infty(c)$
     at time-step $t=10^3$ and related to item $c$ with highest
     $\vone^\top\vK_t(c)$  in the case $N=3$,
     $\Gamma$ defined as in~\eqref{eq:elements_Gamma_mean-field}
     with $\phi=0.75$ and $\iota_{\Gamma}=0.8$, and
     $W$ defined as
     in~\eqref{eq:elements_W_mean-field} with $\iota_W=0.8$. 
     Number of simulations $S=200$
     (sorted on the $x$-axis according to the value of $\vone^\top\vK_t(c)$).
     The crosses represent the value (reported on the $y$-axis) of
     $\widetilde{P}_\infty(c)$ estimated by
     $\vone^\top\vK_{t_\infty}(c)/(Nt_\infty)$ at time-step $t_{\infty}=10^5$.
   }
   \label{Figure:confidence_intervals_mean_field}
  \end{center}
\end{figure}

As an example,  we have simulated $S=200$ independent
  triplets of 
  innovation processes with $\Gamma$ and $W$ of the mean-field type
  defined with $N=3$, $\gamma^{*}=0.75$,
$\iota_{\Gamma}=\iota_{W}=0.8$,
and we show the
confidence intervals~\eqref{eq:confidence_interval_mean_field_case}
of level $(1-\alpha)=95\%$  at time-step $t=10^3$, 
associated to the item $c$ with highest
$\vone^\top\vK_t(c)$ 
(see Figure~\ref{Figure:confidence_intervals_mean_field}).
In this case, the empirical coverage is equal to $0.970$,  
which is again very close to the nominal value $0.95$.

\section{Applications}\label{experiments}

In this section we apply the proposed inferential methodology on the two real data sets,
already considered  in \cite{ale-cri-ghi-innovation-2023},
to test the minimum level of interaction presented by the matrices
$\Gamma$ and $W$ in these specific contexts. 
In particular, the first one is taken from the social content aggregation website Reddit, collected,
elaborated
and made freely available on the web by the authors of~\cite{monti}
(see \url{https://github.com/corradomonti/demographic-homophily}),
while the second one is downloaded from the on-line library
Project Gutenberg, which is a collection of public domain books
(see \url{https://www.gutenberg.org}). 

\subsection{Reddit data set}
The data set includes a collection of news posts and their corresponding comments
from the Reddit community r/news over the period $2016$--$2020$,
downloaded from the website Reddit \cite{monti}.  
Each news item is linked to the user who posted it.
In addition, the data provide the topic associated with each news post
and assign to every comment a sentiment score,
expressed as a real number in the interval $(-1, 1)$.
We restrict our analysis to comments on news classified under the topic
``Politics''.  
The sentiment variable is then discretized as follows:
a comment is labeled as ``positive'' if its sentiment score exceeds $+0.35$,
and as ``negative'' if it is below $-0.35$.
Comments with original sentiment values in $(-0.35, +0.35)$ are excluded.  
The resulting data set contains
$2\,602\,173$ comments in each of the two categories.
Consequently, we define two innovation processes:
one associated with authors receiving at least one negative comment on their posts,
and another associated with authors receiving at least one positive  comment.
A first analysis of this data set can be found in
\cite{ale-cri-ghi-innovation-2023}, where  we illustrate that these
two sequences exhibit empirical behaviours that are
in accordance with those predicted by Theorems~\ref{th-synchro-rates}
and~\ref{th-synchro-probab},   
and we provide the estimates 
$\widehat{\gamma}^{*}=0.781$ and $\widehat{r}=10^{-0.727}=0.187$ 
(these estimates have been obtained by means of a suitable regression on
$(\vD^*_t)_t$, which simulations have shown to be a highly accurate estimation method). 
 We here continue the study  
performing the two hypothesis tests on $\Gamma$ (taking the above estimates as the known values of $\gamma^*$ and $r$) and on $W$
described in Subsection~\ref{sec:case-N2}. 
More precisely, we consider the two one-sided tests:
$$
H_{\Gamma,0}:\;\iota\geq \iota_{\Gamma,0},\,\eta\leq 1/2
\qquad\mbox{and}\qquad
H_{W,0}:\; \iota\geq \iota_{W,0},
$$
where $\iota_{\Gamma,0}$
must be selected in the interval
 $(0.141, \, 0.291]$ and  $\iota_{W,0}$
must be selected in the interval
$(\tfrac{1}{2},\, 2)$.
The $p$-values associated to the null hypothesis
$H_{\Gamma,0}$ and $H_{W,0}$, for different values of $\iota_{\Gamma,0}$ and 
$\iota_{W,0}$, respectively, are collected in Table~\ref{Table Reddit}.
In order to better describe the empirical distribution of the
$p$-values for testing $W$, instead of taking a single
item $c$, we take the $100$ items $c$
with highest $(K_{t,1}(c)+K_{t,2}(c))$, and report in Table~\ref{Table Reddit}
the summary statistics of this sample
(i.e. the quantiles of order $50\%$, $75\%$ and $95\%$).
We can see that, considering a test on $\Gamma$ performed at level
$\alpha=0.05$,
we can reject the null hypothesis with 
$\iota_{\Gamma,0}=0.21$, but we cannot reject it for $\iota_{\Gamma,0}=0.18$.
Regarding the test on $W$ instead, we observe that the proportion of
items $c$ providing a $p-$value that
let us reject the null hypothesis that the interaction
intensity $\iota_W$ is higher than $\iota_{W,0}$ is from
$75\%$ to $95\%$ for $\iota_{W,0}\leq 1.25$, while is even higher
than $95\%$ or $\iota_{W,0}=1.50$.\\

\renewcommand{\arraystretch}{1.5}
\begin{table}[ht!]
\caption{Reddit data set. $p$-values associated to the hypothesis tests
           described in~Subsection~\ref{sec:case-N2} with
      $H_{\Gamma,0}: \iota_{\Gamma}\geq \iota_{\Gamma,0},\,\eta \leq 1/2$ and
      $H_{W,0}: \iota_{W}\geq \iota_{W,0}$, respectively. 
      The table also contains the summary statistics related to the sample 
      of the $100$ items $c$ with highest $(K_{t,1}(c)+K_{t,2}(c))$.}
    \label{Table Reddit}
\centering
    \begin{tabular}{|r|c||r|c|}
     \hline
Interaction Intensity $\iota_{\Gamma,0}$ & 
 p-value &
 Interaction Intensity  $\iota_{W,0}$ & 
 \shortstack{p-value distribution\\ 
$q_{50\%}$,\ \ $q_{75\%}$,\ \ $q_{95\%}$}   \\
 \hline
0.15 & 0.384 & 0.60 & $<$0.001, 0.007, 0.486\\
0.18 & 0.064 & 0.80 & $<$0.001, $<$0.001, 0.229\\
0.21 & 0.014 & 1.00 & $<$0.001, $<$0.001, 0.121\\
0.24 & 0.003 & 1.25 & $<$0.001, $<$0.001, 0.058\\
0.27 & 0.001 & 1.50 & $<$0.001, $<$0.001, 0.029\\
 \hline
    \end{tabular}
\end{table}

\subsection{Gutenberg data set}
This data set is drawn from the on-line library Project Gutenberg
and concerns seven books written in English that belong to two specific
literary genres: ``Western'' and ``History''.  
More precisely, the corpus contains $476\,948$ words for each genre.
The two resulting word sequences constitute the innovation processes under investigation.
A word is classified as new (respectively, old) for the overall system
if it has not (respectively, has already) appeared in any of the considered genres.
A first analysis of also this
data set can be found in \cite{ale-cri-ghi-innovation-2023}, where we show
that the behaviors of these two sequences
predicted by Theorems~\ref{th-synchro-rates}
and~\ref{th-synchro-probab} match
with the observed ones,  
and  we provide the estimates 
$\widehat{\gamma}^{*}=0.466$ and $\widehat{r}=10^{-0.238}=0.578$.  
As in the previous application,
we here continue the study performing
the two hypothesis tests on $\Gamma$ and $W$
described in Subsection~\ref{sec:case-N2}.  
More precisely, we consider the two one-sided tests:
$$
H_{\Gamma,0}:\;\iota\geq \iota_0,\, \eta\leq 1/2
\qquad\mbox{and}\qquad
H_{W,0}:\;\iota\geq \iota_0,
$$
where $\iota_{\Gamma,0}$
must be selected in the interval
$(0.202, \, 0.539]$ and  $\iota_{W,0}$
must be selected in the interval
$(\tfrac{1}{2},\, 2)$.
The $p$-values associated to the null hypothesis
$H_{\Gamma,0}$ and $H_{W,0}$, for different values of $\iota_{\Gamma,0}$ and 
$\iota_{W,0}$, respectively, are collected in Table~\ref{Table Gutenberg}. 
Moreover, we take the $100$ items $c$ 
with highest $(K_{t,1}(c)+K_{t,2}(c))$ and report 
in the table 
the empirical quantiles of the corresponding sample of $p$-values. 
We can observe that, considering a test on $\Gamma$ performed at level
$\alpha=0.05$, we can never reject the null hypothesis.
Regarding the test on $W$ instead, we observe that
the proportion of items $c$ providing a $p$-value
that let us reject the null hypothesis
that the interaction intensity $\iota_W$ is higher than
$\iota_{W,0}$ is from $50\%$ to $75\%$ for $\iota_{W,0}=0.60$,
while is from $75\%$ to $95\%$ for $\iota_{W,0}\geq 0.80$.\\

\renewcommand{\arraystretch}{1.5}
\begin{table}[ht!]
\caption{Gutenberg data set. $p$-values associated to the hypothesis tests
       described in~Subsection~\ref{sec:case-N2} with
      $H_{\Gamma,0}: \iota_{\Gamma}\geq \iota_{\Gamma,0},\, \eta\leq 1/2$ and 
      $H_{W,0}: \iota_{W}\geq \iota_{W,0}$, respectively.
      The table also contains the summary statistics related to the sample 
      of the $100$ items $c$ with highest $(K_{t,1}(c)+K_{t,2}(c))$.}
    \label{Table Gutenberg}
\centering
    \begin{tabular}{|r|c||r|c|}
     \hline
Interaction Intensity $\iota_{\Gamma,0}$ & 
 p-value &
 Interaction Intensity  $\iota_{W,0}$ & 
 \shortstack{p-value distribution\\ 
$q_{50\%}$,\ \ $q_{75\%}$,\ \ $q_{95\%}$}   \\
 \hline
0.21 & 0.034 & 0.60 & 0.010, 0.209, 0.823\\
0.24 & $<$0.001 & 0.80 & $<$0.001, 0.030, 0.699\\
0.27 & $<$0.001 & 1.00 & $<$0.001, 0.005, 0.618\\
0.30 & $<$0.001 & 1.25 & $<$0.001, 0.001, 0.541\\
0.40 & $<$0.001 & 1.50 & $<$0.001, $<$0.001, 0.480\\
 \hline
    \end{tabular}
\end{table}

\section*{Author contributions statement}
All the authors contributed equally to the present work.

\appendix

\renewcommand{\thesection}{\Alph{section}} 
\makeatletter
\renewcommand\@seccntformat[1]{\appendixname\ \csname the#1\endcsname.\hspace{0.5em}}
\makeatother

\paragraph{Notation}
In all the appendix, the notation $R_t=O(s_t)$ means that 
$|R_t|\leq C s_t$ for a suitable deterministic constant $C$ and for $t$ large enough. Similarly, 
the vectorial notation $\vR_t=O(s_t)\vone$ means that each component of $\vR_t$, say $R_{t,j}$, is 
such that $|R_{t,j}|\leq C s_t$ for an appropriate deterministic constant $C$ and for $t$ large enough. 
(Analogous results can be obtained for a suitable integrable random variable $C$.)

\medskip

The results in appendix are given in terms of stable convergence or its variants. 
For more details on these concepts, we refer the reader to~\cite{Ald-Eag-1978, crimaldi-libro, CriLetPra, HallHeyde, Ren}.

\section{General results}

Consider two multi-dimensional real stochastic processes $(\vA_t, \vB_t)_t$, adapted to a filtration $(\mathcal{F}_t)_t$, with the following joint dynamics:
 \begin{equation}\label{eq-general-dynamics}
\begin{split}
\vA_{t+1}&=\vA_t - \frac{1}{t+1}(\phi^{*} I-\Phi^\top)\vA_t + \frac{1}{t+1} \Phi^\top \Delta {\vM}_{t+1} + \vR_{A,t+1},\\
\vB_{t+1}&=\vB_t - \frac{1}{t+1} (\phi^{*}\vB_t-\vA_t) + \frac{1}{t+1} \Delta {\vM}_{t+1} + \vR_{B,t+1}\,, \\   
\end{split}
\end{equation}
where  $\vA_0$ and $\vB_0$ are integrable and 
\begin{itemize}
\item[$(i)$] $\Phi^\top$ is a non-negative irreducible diagonalizable matrix with leading eigenvalue $0<\phi^{*}\leq 1$;
\item[$(ii)$] $\vR_{A,t+1}=O(t^{-(1+\beta)})\vone$ and $\vR_{B,t+1}=O(t^{-(1+\beta)})\vone$
for some $\beta > \frac{\phi^{*}}{2}$;
\item[$(iii)$] $\Delta \vM_{t+1}= O(t^{1-\eta})\vone$ is a martingale difference with $\eta>\phi^{*}/2$ and 
$t^{-(1-\phi^{*})}E[\Delta{\vM}_{t+1}\Delta{\vM}_{t+1}^\top|\mathcal{F}_t]\stackrel{a.s.}\rightarrow  \Sigma_\infty$ 
with $\Sigma_\infty$ a positive semidefinite symmetric {\em random} matrix.
\end{itemize}
It is worthwhile to note that dynamics similar to
\eqref{eq-general-dynamics} have already been considered in the Stochastic Approximation literature. However, the most part of those results do not apply to our situation because the  
covariance matrices in those results are deterministic (e.g.~\cite{laruelle-pages-2013, mok-pel2006}), while the
asymptotic covariance matrices in our CLTs are random. This is also why we do not use the simple convergence in distribution, but we
employ the notion of stable convergence. On the other hand, the paper~\cite{Zhang2016} also takes into account random asymptotic covariance matrices, but it is able to catch our CLT only in the case  
$\phi^*=1$ (compare our assumption $(iii)$ by assumption (2.4) in \cite{Zhang2016}).\\

Let $\vu$ and $\vv$ be the right and the left eigenvectors of $\Phi^\top$ associated to $\phi^*$ and such that
$
\vv^\top\vone=1\quad \mbox{and}\quad \vv^\top\vu=1.
$ (Recall that the components of
these vectors are all strictly positive according to the Frobenius-Perron theory.)
\\
\indent Let $\widetilde{A}_t=\vv^\top \vA_t$ and get its dynamics by multypling $\vv^\top$ to the dynamics of $\vA_t$ as follows:
$$
\widetilde{A}_{t+1} = \widetilde{A}_{t} + \frac{1}{t+1}\phi^{*}\vv^\top \Delta {\vM}_{t+1} + \vv^\top{\vR}_{A,t+1}.
$$
Define
\begin{equation}\label{definitions-hat}
\widehat{\vA}_t=\vA_t-\widetilde{A}_t\vu = (I-\vu\vv^\top)\vA_t 
\qquad\mbox{and}\qquad
\widehat{\vB}_t=\vB_t-\frac{1}{\phi^{*}}\widetilde{A}_t\vu.
\end{equation}
An intuitive reasoning of this decomposition relates to the eigen-structure of the matrix $(\phi^{*} I-\Phi^\top)$ in \eqref{eq-general-dynamics}. Indeed, $\widetilde{A}_t$ is the projection along its kernel, that corresponds to the Perron Frobenius eigen-space of $\Phi$, and $\widehat{\vA}_t$ completes with the projections on the remaining directions.
The process $(\vB_t)_t$ in \eqref{eq-general-dynamics} will follow the dynamics of the attracting process $(\vA_t)_t$, and $\widehat{\vB}_t$ plays the role of $\widehat{\vA}_t$.

\begin{thm}[{CLT for $(\widetilde{A}_t)$ when $N\geq 1$}]\label{thm:TCL_A_tilde} 
In addition to assumptions $(i)$, $(ii)$ and $(iii)$, also assume
\begin{itemize}
\item[$(iv)$] $(\widetilde{A}_t)_t$ is uniformly integrable.
\end{itemize}
Then $(\widetilde{A}_t)_t$ converges almost surely and in mean to an integrable random variable $\widetilde{A}_\infty$ and we have 
$$
t^{\phi^*/2}(\widetilde{A}_t-\widetilde{A}_\infty)\stackrel{stably}\longrightarrow \mathcal{N}(0,\phi^*\vv^\top\Sigma_\infty\vv)\,,
$$
Moreover, the above convergence is in the sense of the almost sure conditional convergence with respect to 
the filtration $(\mathcal{F}_t)_t$.
\end{thm}

\begin{proof}
Set $\breve{M}_t=\sum_{n=1}^t\frac{1}{n}\vv^\top\Delta{\vM}_n$ and $\widetilde{R}_{A,t+1}=\vv^\top{\vR}_{A,t+1}$. 
Therefore, we have 
\begin{equation*}
\begin{split}
\widetilde{A}_{t+1}-\widetilde{A}_0&=
\sum_{n=0}^t \widetilde{A}_{n+1}-\widetilde{A}_{n}
\\
&=
\phi^*\sum_{n=0}^t\frac{1}{n+1}\vv^\top\Delta{\vM}_{n+1}
+\sum_{n=0}^t \widetilde{R}_{A,n+1}\\
&=\phi^*\breve{M}_{t+1}+\sum_{n=0}^t \widetilde{R}_{A,n+1}\,.
\end{split}
\end{equation*}
Since assumption $(iv)$ and $\sum_{n\geq 0} \widetilde{R}_{A,n+1}=\sum_{n\geq 0} O(1/n^{1+\beta})<+\infty$, 
we obtain that $\breve{M}_t\stackrel{a.s./in\, mean}\to \breve{M}_\infty$ for a certain integrable random variable $\breve{M}_\infty$ and so 
$\widetilde{A}_{t}\stackrel{a.s./in\, mean}\to \widetilde{A}_\infty$.
Moreover, we have 
\begin{equation*}
\begin{split}
t^{\phi^*/2}\left(\widetilde{A}_t-\widetilde{A}_\infty\right)
&=
t^{\phi^*/2}\sum_{n\geq t} \left(\widetilde{A}_n-\widetilde{A}_{n+1}\right)\\
&=
t^{\phi^*/2}\phi^*(\breve{M}_{t}-\breve{M}_\infty)+
t^{\phi^*/2}\sum_{n\geq t}\widetilde{R}_{A,n+1}\\
&=t^{\phi^*/2}\phi^*(\breve{M}_{t}-\breve{M}_\infty)+
t^{\phi^*/2}\sum_{n\geq t}O(1/n^{1+\beta})\\
&=t^{\phi^*/2}\phi^*(\breve{M}_{t}-\breve{M}_\infty)+
t^{\phi^*/2}O(1/t^{\beta})\,,
\end{split}
\end{equation*}
where $\beta-\phi^*/2>0$. Therefore, it is enough to prove the convergence of $t^{\phi^*/2}(\breve{M}_{t}-\breve{M}_\infty)$.  
For this purpose, we want to apply Theorem \ref{fam_tri_vet_as_inf}.  Let us consider, for each 
$t$ the filtration $({\mathcal F}_{t,k})_k$ and the process
$(L_{t,k})_k$ defined by
\begin{equation*}
{\mathcal F}_{t,0}={\mathcal F}_{t,1}={\mathcal F}_t, \qquad
L_{t,0}=L_{t,1}=0
\end{equation*}
and, for $k\geq 2$,
\begin{equation*}
{\mathcal F}_{t,k}={\mathcal F}_{t+k-1}, \qquad
L_{t,k}=t^{\phi^*/2}(\breve{M}_t-\breve{M}_{t+k-1}).
\end{equation*}
The process $(L_{t,k})_k$ is a martingale w.r.t. $({\mathcal F}_{t,k})_k$ 
which converges (for $k\to +\infty$) a.s. and in mean to the random
variable $L_{t,\infty}=t^{\phi^*/2}(\breve{M}_t-\breve{M}_{\infty})$. 
In addition, the increment $Y_{t,n}=L_{t,n}-L_{t,n-1}$ is equal to zero
for $n=1$ and, for $n\geq 2$, it coincides with a random variable of
the form $t^{\phi^*/2}(\breve{M}_k-\breve{M}_{k+1})$ with $k\geq
t$. Therefore, we have
\begin{equation*}
\begin{split}
\sum_{n\geq 1} Y_{t,n}^2
&=
t^{\phi^*}\sum_{k\geq t}(\breve{M}_{k}-\breve{M}_{k+1})^2\\
&=
t^{\phi^*}\sum_{k\geq t}\frac{1}{(k+1)^2}
(\vv^\top\Delta{\vM}_{k+1})^2\\
&=
t^{\phi^*}\sum_{k\geq t}\frac{k^{1-\phi^*}}{(k+1)^2}
k^{-(1-\phi^*)}(\vv^\top\Delta{\vM}_{k+1})^2
\stackrel{a.s.}{\longrightarrow}
\vv^\top\Sigma_\infty\vv/\phi^*\,,
\end{split}
\end{equation*}
where the almost sure convergence follows by applying \cite[Lemma 4.1]{cri-dai-min}
and by the fact that 
$$
k^{-(1-\phi^*)}E[(\vv^\top\Delta{\vM}_{k+1})^2|\mathcal{F}_t]=
k^{-(1-\phi^*)}\vv^\top E[\Delta{\vM}_{k+1}\Delta\vM_{k+1}^\top|\mathcal{F}_t]\vv
\stackrel{a.s.}\longrightarrow
\vv^\top\Sigma_\infty\vv\,.
$$
Finally, we have
$$
\sup_{n\geq 1}|Y_{t,n}|=
t^{\phi^*/2}\, \sup_{k\geq t} | \breve{M}_k-\breve{M}_{k+1}|
\leq \sup_{k\geq t}\, k^{\phi^*/2} \frac{|\vv^\top\Delta\vM_{k+1}|}{(k+1)} 
$$
and, since $\Delta \vM_{k+1}= O((k+1)^{1-\eta})\vone$ by assumption $(iii)$ and so 
$k^{\phi^*/2} \frac{|\vv^\top\Delta\vM_{k+1}|}{(k+1)}=O(k^{-(\eta-\phi^*/2)})$,
the thesis follows by Theorem \ref{fam_tri_vet_as_inf}.
\end{proof}

\begin{thm}[{Convergence result for
  $(\widehat{A}_t,\widehat{B}_t)$ when $N=1$}]\label{thm:CLT_N=1} 
Under the assumptions $(i)$ and $(ii)$, we have 
$$\widehat{A}_t\equiv 0
\qquad\mbox{and}\qquad
\widehat{B}_t=B_t-\frac{1}{\phi^{*}} A_t \stackrel{a.s.}= o(t^{-\frac{\phi^{*}}{2}}).$$
\end{thm}

\begin{proof}
First, when $N=1$ it is immediate to see $\widetilde{A}_t\equiv A_t$ and hence $\widehat{A}_t=A_t-\widetilde{A}_t\equiv 0$ for any $t\geq 1$.
The dynamics of $\widehat{B}_t$ can be obtained by subtracting the dynamics of 
$\frac{1}{\phi^{*}}\widetilde{A}_t=\frac{1}{\phi^{*}}A_t$ to the dynamics of $B_t$ as follows:
$$
\widehat{B}_{t+1}=\widehat{B}_t - \frac{1}{t+1} (\phi^{*}\widehat{B}_t+\phi^{*}\frac{1}{\phi^{*}}\widetilde{A}_t-A_t)
+\frac{1}{t+1} \Delta {M}_{t+1} 
- \frac{1}{\phi^{*}}(\widetilde{A}_{t+1}-\widetilde{A}_t) + {R}_{B,t+1}\,,
$$
from which we can derive
$$
\widehat{B}_{t+1}=\widehat{B}_t - \frac{1}{t+1} (\phi^{*}\widehat{B}_t)
+\frac{1}{t+1} \Delta {M}_{t+1} 
- \frac{1}{\phi^{*}}\left(\frac{1}{t+1}\phi^{*}\Delta {M}_{t+1} \right) + {R}_{B,t+1} - \frac{1}{\phi^{*}}{R}_{A,t+1}\,,
$$
that is
$$
\widehat{B}_{t+1}=\left(1 - \frac{\phi^{*}}{t+1}\right)\widehat{B}_t
 + O(t^{-(1+\beta)}).
$$
Then, since $\sum_t\frac{\phi^{*}}{t+1}=+\infty$ and 
$\sum_tO(t^{-(1+\beta)})<+\infty$, we can apply Lemma \ref{lemmaS2.2} to obtain $\widehat{B}_{t}\stackrel{a.s.}\rightarrow 0$.
Moreover, if we multiply the previous dynamics of $\widehat{B}_t$ by $(t+1)^{\frac{\phi^{*}}{2}}$ and 
we use the relation 
$(\frac{t+1}{t})^{\frac{\phi^{*}}{2}}=(1+\frac{1}{t})^{\frac{\phi^{*}}{2}}=1+\frac{\phi^{*}/2}{t+1}+O(1/t^{2})$, 
we get the following dynamics for the process $\widehat{B}^{*}_{t}=t^{\frac{\phi^{*}}{2}}\widehat{B}_{t}$:
$$
\widehat{B}^{*}_{t+1}=
\left(1 - \frac{\phi^{*}}{t+1}\right)
\left(1 + \frac{\phi^{*}/2}{t+1}\right)\widehat{B}_{t}^{*}
 + O(t^{-(1+\beta-\frac{\phi^{*}}{2})}) +
 O(t^{-(2-\frac{\phi^{*}}{2})})\,,
$$
which is equivalent to
$$
\widehat{B}^{*}_{t+1}=
\left(1 - \frac{\phi^{*}/2}{t+1}\right)\widehat{B}_{t}^{*}
 + O(t^{-(1+\beta-\frac{\phi^{*}}{2})})+
 O(t^{-(2-\frac{\phi^{*}}{2})}).
$$
Therefore, since $\sum_t t^{-(2-\frac{\phi^{*}}{2})} < +\infty$ because $\phi^*/2<1$ and
 $\sum_t t^{-(1+\beta-\frac{\phi^{*}}{2})} < +\infty$ because $\beta>\phi^{*}/2$ by assumption, 
we can apply Lemma \ref{lemmaS2.2} also to the 
sequence $\widehat{B}^{*}_{t}$ to obtain
$\widehat{B}^{*}_{t}\stackrel{a.s.}\rightarrow 0$, i.e. $\widehat{B}_{t}\stackrel{a.s.}=o(t^{-\frac{\phi^{*}}{2}})$.
\end{proof}

We recall that the diagonalizable and irreducible matrix $\Phi^\top$ can be decomposed as 
\begin{equation}\label{eq-def-D}
\Phi^\top=\phi^{*}\vu\vv^\top+UDV^\top\,,
\end{equation}
where $D$ is the diagonal matrix whose elements
are the eigenvalues $\phi_j$, $j\geq 2$ of $\Phi$ (i.e. of $\Phi^\top$) different from $\phi^*$ and 
 $U$ and $V$ denote the matrices whose columns are
the left (right) and the right (left) eigenvectors of $\Phi$ (of $\Phi^\top$, respectively) associated to these eigenvalues, such that we have 
\begin{equation}\label{matrices-prop}
  V^\top\vu=U^\top\vv=0,\quad V^\top U=U^\top V=I,\quad
  I=\vu\vv^\top + UV^\top\,.
\end{equation}
Let $\phi_2^{*}$ be an eigenvalue of $\Phi$ different from $\phi^*$ with highest real part, 
i.e. $\phi_2^{*}\in Sp(\Phi)\setminus\{\phi^{*}\}$ such that 
$\mathcal{R}e(\phi_2^{*})=\max\{\mathcal{R}e(\phi):\, \phi\in Sp(\Phi)\setminus\{\phi^{*}\}\}$.

\begin{thm}[{CLT for $(\widehat{\vA}_t,\widehat{\vB}_t)$ when $N\geq 2$}]\label{clt-hat-N}
 In addition to assumptions $(i)$, $(ii)$ and $(iii)$, assume 
\begin{itemize}
\item[$(v)$] ${\mathcal Re}(\phi_2^{*})/\phi^{*}<1/{2}$.
\end{itemize}
Then, we have
$$
t^{\phi^*/2}(\widehat{\vA}_t,\widehat{\vB}_t)\stackrel{stably}\longrightarrow 
\mathcal{N}(\vzero, {\mathcal M}_\infty)
$$
with 
\begin{equation}\label{eq-limit-matrix}
\mathcal{M}_\infty=\begin{pmatrix}
{\mathcal M}_\infty^{11} &{\mathcal M}_\infty^{13}\\
{\mathcal M}_\infty^{13} &{\mathcal M}_\infty^{33}
\end{pmatrix}=
\begin{pmatrix}
U{\mathcal S}_\infty^{11}U^\top &U{\mathcal S}_\infty^{13}U^\top\\
U{\mathcal S}_\infty^{13}U^\top &U{\mathcal S}_\infty^{33}U^\top
\end{pmatrix}\,,
\end{equation}
where, 
 for $h,j=2,\dots, N$
\begin{equation*}
\begin{split}
[{\mathcal S}_\infty^{11}]_{h,j}&=\frac{\phi_j\phi_h}{\phi^*-\phi_j-\phi_h}(\vv_j^\top \Sigma_\infty\vv_h),\\
[{\mathcal S}_\infty^{33}]_{h,j}&=\frac{1}{\phi^*-\phi_j-\phi_h}(\vv_j^\top \Sigma_\infty\vv_h),\\
[{\mathcal S}_\infty^{13}]_{h,j}&=\frac{\phi_j}{\phi^*-\phi_j-\phi_h}(\vv_j^\top \Sigma_\infty\vv_h).
\end{split}
\end{equation*}
\end{thm}

\begin{proof} We split the proofs in some steps.
\\

\indent {\em Step 1 (Dynamics of $\theta_t=(\widehat{A}_t,\widehat{B}_t)$).}\\
\noindent 
First of all, we observe that $\widehat{\vA}_t=UV^\top\vA_t$ and  
the dynamics of $\widehat{\vA}_t$ can be obtained by multypling by $UV^\top$ the dynamics of $\vA_t$ as follows:
$$
\widehat{\vA}_{t+1}=\widehat{\vA}_t - \frac{1}{t+1}U(I\phi^{*}-D)V^\top\widehat{\vA}_t + 
\frac{1}{t+1} UDV^\top \Delta {\vM}_{t+1} + UV^\top{\vR}_{A,t+1}.
$$
The dynamics of $\widehat{\vB}_t$ can be obtained by subtracting the dynamics of $\frac{1}{\phi^{*}}\widetilde{A}_t\vu$ 
to the dynamics of $\vB_t$ as follows:
$$
\widehat{\vB}_{t+1}=\widehat{\vB}_t - \frac{1}{t+1} (\phi^{*}\widehat{\vB}_t+\phi^{*}\frac{1}{\phi^{*}}\widetilde{A}_t\vu-\widehat{\vA}_t-\widetilde{A}_t\vu)
+\frac{1}{t+1} \Delta {\vM}_{t+1} 
- \frac{1}{\phi^{*}}\vu(\widetilde{A}_{t+1}-\widetilde{A}_t) + {\vR}_{B,t+1}\,,
$$
from which we can derive
$$
\widehat{\vB}_{t+1}=\widehat{\vB}_t - \frac{1}{t+1} (\phi^{*}\widehat{\vB}_t-\widehat{\vA}_t)
+ \frac{1}{t+1} (I-\vu\vv^\top)\Delta {\vM}_{t+1} 
 + {\vR}_{B,t+1}
 - \frac{1}{\phi^{*}}\vu\vv^\top{\vR}_{A,t+1}\,.
$$
Then the dynamics of $\veta_t=(\widehat{\vA}_t,\widehat{\vB}_t)$ can be expressed as
$$
\veta_{t+1}-\veta_t =
-\frac{1}{t+1}Q\veta_t + \frac{1}{t+1}R\Delta\vM_{\veta,t+1} + \vR_{\veta,t+1}\,,
$$
where\footnote{The symbol $\mathcal{O}$ denotes the null matrix.}
$$
Q = \begin{pmatrix}
 U(I\phi^{*}-D)V^\top & \mathcal{O}\\
 -UV^{\top} & I\phi^{*}
\end{pmatrix},
\qquad
\Delta\vM_{\veta,t+1} = \begin{pmatrix} 
\Delta\vM_{t+1}\\
\Delta\vM_{t+1}
\end{pmatrix},
\qquad
R = \begin{pmatrix}
 UDV^\top & \mathcal{O}\\
 \mathcal{O} & UV^{\top}
\end{pmatrix}.
$$
Now, let us define the $(2N)\times(2N-1)$ matrices
$$
U_\veta=
\begin{pmatrix}
 U & \vzero & 0 \\
 0 & \vu & U \\
\end{pmatrix},
\qquad\mbox{and}\qquad
V_\veta^{\top}=
\begin{pmatrix}
 V^{\top} & 0 \\
 \vzero^{\top} & \vv^{\top}\\
 0 & V^{\top} \\
\end{pmatrix},
$$
so that $V_\veta^{\top}U_\veta=I$ and 
$$
U_\veta V_\veta^{\top} =
 \begin{pmatrix}
 UV^{\top} & 0 \\
 \vzero^{\top} & I
\end{pmatrix}.
$$
Then, defining the $(2N-1)\times(2N-1)$ matrices
$$
S_Q=\begin{pmatrix}
 I\phi^{*}- D & \vzero & 0\\
 \vzero^\top & \phi^{*} & \vzero^\top\\
 -I & \vzero & I\phi^{*}
\end{pmatrix},
\qquad\mbox{and}\qquad
S_R=\begin{pmatrix}
 D & \vzero & 0\\
 \vzero^\top & 0 & \vzero^\top\\
 0 & \vzero & I
\end{pmatrix},
$$
we have that $Q=U_\veta S_Q V_\veta^{\top}$ and  $R=U_\veta S_R V_\veta^{\top}$.
Then, from the above relations on $U_\veta$ and $V_\veta^{\top}$ we get $U_\veta V_\veta^{\top}\veta_t=\veta_t$, and so
multypling the dynamics of $\veta_t$ by $U_\veta V_\veta^{\top}$ we finally get
$$
\veta_{t+1} = \frac{1}{t+1}U_\veta\left(I-\frac{1}{t+1}S_Q\right)
V_\veta^\top\veta_t + \frac{1}{t+1}R\Delta\vM_{\veta,t+1} + \vR_{\veta,t+1}.
$$

\indent {\em Step 2 (Explicit expression for $\veta_t=(\widehat{\vA}_t,\widehat{\vB}_t)$).}
\\
\noindent Recalling that the eigenvalues of $\Phi^\top$ different from $\phi^*$ are denoted by
$\phi_j$ for $j=2,\dots,N$, let us   
set $\alpha_j=\phi^{*}-\phi_j$. Recall that $\mathcal{R}e(\alpha_j)>0$ since $\phi^*$ is the leading eigenvalue and so 
$\mathcal{R}e(\phi_j)<\phi^{*}$.
Then, setting an initial time $m_0$
such that ${\mathcal Re}(\alpha_j)/(m_0+1)<1$ for each $j$,
 we can write
$$
\begin{aligned}
\veta_{t+1} &&=&\ C_{m_0,t}\veta_{m_0} + 
\sum_{k=m_0}^t\frac{1}{k+1}C_{k+1,t}R\Delta\vM_{\veta,k+1} +
\sum_{k=m_0}^t C_{k+1,t} \vR_{\veta,k+1} \\
&&=&\ 
C_{m_0,t}\veta_{m_0} + 
\sum_{k=m_0}^t\vT_{t,k+1} + \boldsymbol{{\rho}}_{t+1}\,,
\end{aligned}
$$
where
\begin{equation*}\label{def:A_k_t}
\begin{split}
C_{k+1,t} &= U_\veta A_{k+1,t} V_\veta^{\top},
\qquad\mbox{and}\\
A_{k+1,t} &= 
\begin{cases}
\prod_{m=k+1}^t \left(I-\frac{1}{m+1}S_Q\right)
=
\begin{pmatrix}
A^{11}_{k+1,t} & \vzero & 0\\
0 & a^{22}_{k+1,t} & 0\\
A^{31}_{k+1,t} & \vzero & A^{33}_{k+1,t}\\
\end{pmatrix}
\quad & k=m_0-1,\dots, t-1,\\
I\quad &k=t\,.
\end{cases}
\end{split}
\end{equation*}
Notice that the blocks $A^{11}_{k+1,t}$, $A^{31}_{k+1,t}$ and $A^{33}_{k+1,t}$ are all diagonal $(N-1)\times(N-1)$ matrices. 
Moreover, setting for any $x\in\mathbb{C}$
with ${\mathcal Re}(x)/(m_0+1)<1$, $p_{m_0-1}(x)=1$ and $p_k(x)=\prod_{m=m_0}^{k}(1-\tfrac{x}{m+1})$ for $k\geq m_0$ and 
$F_{k+1,t}(x)=\frac{p_t(x)}{p_k(x)}$ for $m_0-1\leq k\leq t-1$, 
we get (see \cite[Lemma~A.5]{ale-cri-ghi-MEAN} for similar technical calculations)
$$
\begin{aligned}
[A^{11}_{k+1,t}]_{jj} &&=&\ F_{k+1,t}(\alpha_j)\\
[A^{33}_{k+1,t}]_{jj} &&=&\ a^{22}_{k+1,t}\ = \ F_{k+1,t}(\phi^{*})\\
[A^{31}_{k+1,t}]_{jj} &&=&\ \left\{
\begin{aligned}
& -\frac{F_{k+1,t}(\alpha_j)-F_{k+1,t}(\phi^{*})}{\alpha_j-\phi^{*}} & \mbox{for } \alpha_j\neq \phi^{*} \mbox{ ,i.e. } \phi_j\neq 0\\
& \frac{1-\phi^{*}}{\phi^{*}} F_{k+1,t}(\phi^{*}) \ln\left(\frac{t}{k}\right) +
O\left(\frac{1}{t}\right)& \mbox{for } \alpha_j = \phi^{*} \mbox{ ,i.e. } \phi_j=0.
\end{aligned}
\right.
\end{aligned}
$$

\indent {\em Step 3 (Study of the terms $C_{m_0,t}\veta_{m_0}$ and $\boldsymbol{{\rho}}_{t+1}$).}
\\
\noindent We will prove that $t^{\frac{\phi^{*}}{2}}|C_{m_0,t}\veta_{m_0}|\stackrel{a.s.}\to 0$ and 
$t^{\frac{\phi^{*}}{2}}|\boldsymbol{{\rho}}_{t+1}|\stackrel{a.s.}\to 0$. 
To this end, first notice that $O(|C_{k+1,t}|)=O(|A_{k+1,t}|)$ and, setting 
$a_2^*=\phi^*-{\mathcal Re}(\phi^*_2)$, we have that
$$
\begin{aligned}
|A_{k+1,t}| &&=&\ 
O\left(\frac{|p_t(\alpha_2^{*})|}{|p_k(\alpha_2^{*})|}\right) +  
O\left(\frac{|p_t(\phi^{*})|}{|p_k(\phi^{*})|}\right) +
O\left(\frac{|p_t(\phi^{*})|}{|p_k(\phi^{*})|}\ln\left(\frac{t}{k}\right)\right) +
O\left(\frac{1}{t}\right)\\
&&=&\ 
O\left(\left(\frac{k}{t}\right)^{a_2^{*}}\right) + 
O\left(\left(\frac{k}{t}\right)^{\phi^{*}}\right) +
O\left(\left(\frac{k}{t}\right)^{\phi^{*}}\ln\left(\frac{t}{k}\right)\right) + 
O\left(\frac{1}{t}\right)\\
&&=&\ 
O\left(\left(\frac{k}{t}\right)^{a_2^{*}}\right) + 
O\left(\left(\frac{k}{t}\right)^{\phi^{*}}\ln(t)\right)\\
&&=&\ 
O\left(\left(\frac{k}{t}\right)^{(\phi^*-{\mathcal Re}(\phi^*_2)}\right) + 
O\left(\left(\frac{k}{t}\right)^{\phi^{*}}\ln(t)\right)
\quad\mbox{for } k=m_0-1,\dots, t-1,
\end{aligned}
$$
and simply $|A_{t+1,t}|=O(1)$ for $k=t$. 
Therefore, since assumption $(v)$, we have 
$$
  t^{\frac{\phi^{*}}{2}}|C_{m_0,t}\veta_{m_0}|=
  |\veta_{m_0}|\, O\left(
\left(\frac{1}{t}\right)^{\phi^*/2-{\mathcal Re}(\phi_2^{*})} + 
\left(\frac{1}{t}\right)^{\phi^{*}/2}\ln(t)
\right)
\stackrel{a.s.}\longrightarrow 0\,.
$$ 
Moreover, recalling that $\vR_{\veta,t+1}=O(t^{-(1+\beta)})\vone$ for some $\beta > \frac{\phi^{*}}{2}$,  we have
$$
\begin{aligned}
t^{\frac{\phi^{*}}{2}}|\boldsymbol{{\rho}}_{t+1}| &&=&\ t^{\frac{\phi^{*}}{2}}\left|
\sum_{k=m_0}^t  C_{k+1,t} \vR_{\veta,k+1}
\right|\\ 
&&=&\ t^{\frac{\phi^{*}}{2}} O\left( 
\sum_{k=m_0}^{t-1} \left(\frac{k}{t}\right)^{a_2^{*}} \frac{1}{k^{{1+\beta}}} + 
\sum_{k=m_0}^{t-1} \left(\frac{k}{t}\right)^{\phi^{*}}\ln(t)\frac{1}{k^{{1+\beta}}}
\right)+
t^{\phi^*/2}O(1/t^{1+\beta})\\
&&=&\ t^{\frac{\phi^{*}}{2}} O\left( \frac{1}{t^{a_2^{*}}}
\sum_{k=m_0}^{t-1} k^{a_2^{*}-1-\beta} + 
\frac{\ln(t)}{t^{\phi^{*}}}
\sum_{k=m_0}^{t-1} k^{\phi^{*}-1-\beta}
\right)+ 
O(1/t^{1+\beta-\phi^*/2}),
\end{aligned}
$$
which tends to zero because $\beta>\frac{\phi^{*}}{2}$
and $a_2^*>0$.\\

\indent{\em Step 4 (Study of the ``delta-martingale'' term).}
\\
We want to apply Theorem~\ref{th-triangular}. Therefore we have to check conditions (1) and (2) in Theorem~\ref{th-triangular}.
\\
\indent Since the relation $V_\veta^\top U_\veta = I$ implies $V_\veta^\top R = S_R V_\veta^\top$, we have that
$$
\begin{aligned}
\sum_{k=m_0}^t(t^{\frac{\phi^{*}}{2}}\vT_{t,k+1})(t^{\frac{\phi^{*}}{2}}\vT_{t,k+1}^\top)&&=&\ 
t^{\phi^{*}} \sum_{k=m_0}^t\frac{1}{(k+1)^2}C_{k+1,t}R\Delta\vM_{\veta, k+1}\Delta\vM_{\veta, k+1}^\top R^\top C_{k+1,t}^\top\\
&&=&\ U_\veta \left(t^{\phi^{*}} \sum_{k=m_0}^t\frac{1}{(k+1)^2}A_{k+1,t}V_\veta^\top  R\Delta\vM_{\veta, k+1}\Delta\vM_{\veta, k+1}^\top R^\top  V_\veta A^\top_{k+1,t}\right) U_\veta^\top\\
&&=&\ U_\veta \left(t^{\phi^{*}} \sum_{k=m_0}^t\frac{1}{(k+1)^2}A_{k+1,t} S_R V_\veta^\top\Delta\vM_{\veta, k+1}\Delta\vM_{\veta, k+1}^\top 
V_\veta S_R A^\top_{k+1,t}\right) U_\veta^\top.
\end{aligned}
$$
Therefore, omitting the last term of the sum because $|A_{t+1,t}|=O(1)$ and , by assumption $(iii)$, we have 
 $(t^{\phi^*}/(t+1)^2)|\Delta\vM_{t+1}|=O(t^{\phi^*-2+1-\eta})=O(t^{-(1-\phi^*+\eta)})\to 0$, 
 it is enough to study the convergence of 
\begin{equation}\label{eq-sum}
t^{\phi^{*}} \sum_{k=m_0}^{t-1}\frac{1}{(k+1)^2}A_{k+1,t}S_R V_\veta^\top\Delta\vM_{\veta, k+1}\Delta\vM_{\veta, k+1}^\top V_\veta S_R A^\top_{k+1,t}.
\end{equation}
To this purpose, setting 
$B_{\veta, k+1}=V_\veta^\top\Delta\vM_{\veta, k+1}\Delta\vM_{\veta, k+1}^\top V_\veta$, 
$B_{k+1}=V^\top\Delta\vM_{k+1}\Delta\vM_{k+1}^\top V$,  
$\vb_{k+1}=V^\top\Delta\vM_{k+1}\Delta\vM_{k+1}^\top \vv$ 
and $b_{k+1}=\vv^\top\Delta\vM_{k+1}\Delta\vM_{k+1}^\top \vv$, we observe that
$$
B_{\veta,k+1}=
\begin{pmatrix}
B_{k+1} & \vb_{k+1} & B_{k+1}\\
\vb^\top_{k+1} & b_{k+1} & \vb^\top_{k+1}\\
B_{k+1} & \vb_{k+1} & B_{k+1}
\end{pmatrix}.
$$
Since in $B_{\veta,k+1}$ the first and the third row and column of blocks are the same, 
the matrix $(A_{k+1,t}S_R)$ can be replaced by a diagonal matrix, with the following diagonal blocks: 
$A_{k+1,t}^1=A_{k+1,t}^{11}D$,  $A_{k+1,t}^3=A_{k+1,t}^{31}D + A_{k+1,t}^{33}$ and $a_{k+1,t}^{2}=0$.
Hence, the above expression \eqref{eq-sum} can be rewritten as
$$
t^{\phi^{*}} \sum_{k=m_0}^{t-1}\frac{1}{(k+1)^2}
\begin{pmatrix}
A_{k+1,t}^1 B_{k+1}A_{k+1,t}^1 & \vzero & A_{k+1,t}^1 B_{k+1}A_{k+1,t}^3\\
\vzero^\top & 0 & \vzero^\top\\
A_{k+1,t}^3 B_{k+1}A_{k+1,t}^1 & \vzero & A_{k+1,t}^3 B_{k+1}A_{k+1,t}^3
\end{pmatrix}.
$$
The elements of $A_{k+1,t}^1$ and $A_{k+1,t}^3$ in the above
matrix can be rewritten in terms of $F_{k+1,t}(\cdot)$ as 
$$
\begin{aligned}
  [A^{1}_{k+1,t}]_{jj} &&=&\ (\phi^{*}-\alpha_j)F_{k+1,t}(\alpha_j) =
  \phi_jF_{k+1,t}(\alpha_j)\\
[A^{3}_{k+1,t}]_{jj} &&=&\ \left\{
\begin{aligned}
& F_{k+1,t}(\alpha_j) & \mbox{for } \alpha_j\neq \phi^{*}, \mbox{ i.e. } \phi_j\neq 0\\
& F_{k+1,t}(\phi^{*})& \mbox{for } \alpha_j = \phi^{*}, \mbox{ i.e. } \phi_j=0.
\end{aligned}
\right.
\end{aligned}
$$
Moreover, with some computations (see \cite[Lemma~A.3]{ale-cri-ghi-WEIGHT-MEAN} for similar technical calculations) for complex numbers $x$, $y$ such that ${\mathcal Re}(x+y) - \phi^*>0$
we have 
\begin{equation}\label{eq:bounded}
    t^{\phi^{*}}\sum_{k=m_0}^{t-1} \frac{1}{(k+1)^{1+\phi^{*}}}
F_{k+1,t}(x)F_{k+1,t}(y)\rightarrow \frac{1}{x+y-\phi^*}.
\end{equation}
Hence, combining (see \cite[Lemma 4.1]{cri-dai-min}) this limit with the limit relation
$$
t^{-(1-\phi^{*})}E[B_{t+1}|\mathcal{F}_{t}] = 
t^{-(1-\phi^{*})}V^\top E[\Delta\vM_{t+1}\Delta\vM_{t+1}^\top|\mathcal{F}_{t}]V\rightarrow V^\top \Sigma_\infty V, 
$$
we get 
\begin{equation}\label{eq:relation_limit}
t^{\phi^{*}}\sum_{k=m_0}^{t-1} \frac{1}{(k+1)^{1+\phi^{*}}}
\left((k+1)^{-(1-\phi^{*})}B_{k+1}\right)
F_{k+1,t}(x)F_{k+1,t}(y)\rightarrow \frac{1}{x+y-\phi^*} V^\top \Sigma_\infty V\,.
\end{equation}
As a consequence, recalling that $\alpha_j=\phi^*-\phi_j$
and, by assumption $(v)$, $\phi^*-2{\mathcal Re}(\phi^*_2)>0$, the almost sure convergences of all the elements of the previous matrix can be obtained by applying \eqref{eq:relation_limit} as follows:
$$
\begin{aligned}
& t^{\phi^{*}}\sum_{k=m_0}^{t-1}\frac{1}{(k+1)^2}[A_{k+1,t}^1 B_{k+1}A_{k+1,t}^1]_{h,j}\rightarrow 
  [{\mathcal S}_\infty^{11}]_{h,j}=
  \frac{\phi_j\phi_h}{\phi^*-\phi_j-\phi_h}(\vv_j^\top \Sigma_\infty\vv_h),\\
  & t^{\phi^{*}}\sum_{k=m_0}^{t-1}\frac{1}{(k+1)^2}[A_{k+1,t}^3 B_{k+1}A_{k+1,t}^3]_{h,j}
  \rightarrow 
      [{\mathcal S}_\infty^{33}]_{h,j}=
      \frac{1}{\phi^*-\phi_j-\phi_h}(\vv_j^\top \Sigma_\infty\vv_h),\\
      & t^{\phi^{*}}\sum_{k=m_0}^{t-1}\frac{1}{(k+1)^2}[A_{k+1,t}^1 B_{k+1}A_{k+1,t}^3]_{h,j}
      \rightarrow 
          [{\mathcal S}_\infty^{13}]_{h,j}=
          \frac{\phi_j}{\phi^*-\phi_j-\phi_h}(\vv_j^\top \Sigma_\infty\vv_h).
\end{aligned}
$$
Therefore condition (1) of Theorem~\ref{th-triangular} is satisfied with the matrix 
${\mathcal M}_\infty$ defined in \eqref{eq-limit-matrix}.
\\

\indent Regarding condition (2) of Theorem~\ref{th-triangular}, we observe that, by assumption $(iii)$, we have 
\begin{equation*}
\begin{split}
|{\mathbf T}_{t,k+1}|
&=O\left( \frac{1}{k+1} |A_{k+1,t}||\Delta\vM_{\veta,k+1}|\right)
 =O\left(|A_{k+1,t}|(k+1)^{-\eta}\right)\\
&=O\left(k^{-\eta}\,\left(\frac{k}{t}\right)^{(\phi^*-{\mathcal Re}(\phi^*_2)}\right) + 
  O\left(k^{-\eta}\,\left(\frac{k}{t}\right)^{\phi^{*}}\ln(t)\right)
\quad\mbox{for } k=m_0,\dots, t-1
\end{split}
\end{equation*}
and $|{\mathbf T}_{t,t+1}|=O\left( \frac{1}{t+1} |A_{t+1,t}||\Delta\vM_{\veta,t+1}|\right) = O(t^{-\eta})$ for $k=t$. 
Hence, for any $u$, we have 
\begin{align*}
  \left(\sup_{m_0\leq k\leq t}
  \Big|t^{\phi^*/2} {\vT}_{t,k+1}\Big|\right)^{2u} &\leq
  t^{\phi^* u} \sum_{k=m_0}^{t-1}
  |{\vT}_{t,k+1}|^{2u} + t^{\phi^* u}|{\vT}_{t,t+1}|^{2u}
  \\ &= 
  O\left( t^{-(\phi^*-2{\mathcal Re}(\phi^*_2))u} \sum_{k=m_0}^{t-1} k^{-(\eta-\phi^{*}+2{\mathcal Re}(\phi^*_2))u}\right)\\
  &+
  O\left( \frac{\ln(t)^{2u}}{t^{\phi^*u}}
  \sum_{k=m_0}^{t-1} k^{-(\eta-\phi^{*})2u}\right) + 
  O(t^{-(2\eta-\phi^* )u}),
\end{align*}
which tends to zero
for $u>\max\{1/\eta; 1/(2\eta-\phi^*)\}$ 
since $\phi^*-2{\mathcal Re}(\phi^*_2)>0$ and 
$\eta>\phi^{*}/2$.
This result, in particular, implies 
condition $(2)$ of Theorem~\ref{th-triangular}.
\\

\indent By Theorem~\ref{th-triangular}, we can finally conclude that  
$t^{\phi^*/2}\sum_{k=m_0}^{t}{\mathbf T}_{t,k+1}$ converges stably to the
Gaussian kernel ${\mathcal N}(\vzero, {\mathcal M}_\infty)$.
\end{proof}

\begin{rem}\label{suppl:rem:diag-irrid}
Relaxing the assumption of diagonalizability of the matrix $\Phi$ would lead to analogous results. Indeed, it is enough to work with the associated Jordan matrix, 
instead of the matrix $D$ in~\eqref{eq-def-D}. However,  
doing so would substantially increase the notational complexity and the length of the paper, making it unnecessarily cumbersome. Note that the fundamental ideas of the proof will remain the same, only the computations and the final formulas will become more complex and difficult to read. To get an idea, we refer to~\cite{yang2025}, where, for a different context, the assumption of a diagonalizable matrix is replaced by the assumption of a block upper-triangular matrix. Note that the proofs are still 
based on~\cite[Lemma~A.4]{ale-cri-ghi}, \cite[Lemma~A.5]{ale-cri-ghi-MEAN} and its analogous~\cite[Lemma~A.5]{yang2025}.
 Moreover, in the case of a reducible matrix $\Phi$, we refer to~\cite{ale-cri-ghi-innovation-2023, ale-cri-ghi-complete, ale-ghi}.
\end{rem}

\begin{rem}\label{rem:divergence assumption second eigenvalue}
A key point in the proof of Theorem~\ref{clt-hat-N} is the convergence of the left-hand term of \eqref{eq:bounded} toward a bounded limit, which is ensured by assumption (v), i.e. ${\mathcal Re}(\phi_2^{*})<\phi^*/{2}$.  On the other hand, when ${\mathcal Re}(\phi_2^{*})\geq\phi^{*}/2$, this quantity can diverge to $+\infty$, implying the divergence of $t^{\phi^*/2}(\widehat{\vA}_t,\widehat{\vB}_t)$.  
In fact, since $\Phi$ is a real matrix,
in its spectrum we have $\phi_2^{*}$ and its conjugate, so that $F_{k+1,t}(x)F_{k+1,t}(y)=|F_{k+1,t}(x)|^2$ 
when we take $x=\alpha_2^*=\phi^*-\phi_2^{*}$ and for $y$ its conjugate (or $y=x$ if $x$ is real). Now, we observe that $|F_{k+1,t}(\alpha_2^*)|^2$ increases with ${\mathcal Re}(\phi_2^{*})$ and   the corresponding limit in \eqref{eq:bounded}, that is $1/(2{\mathcal Re}(\alpha_2^*)-\phi^*)=1/(\phi^*-2{\mathcal Re}(\phi_2^*))$, goes to $+\infty$ for ${\mathcal Re}(\phi_2^*)\uparrow \phi^{*}/2$. As a consequence of these two facts, 
we can conclude that the limit of the left-hand term of \eqref{eq:bounded} is necessarily $+\infty$ whenever ${\mathcal Re}(\phi_2^{*})\geq\phi^{*}/2$. 
This means that, in that case, the rate of convergence is slower than $t^{\phi^*/2}$.
\end{rem}

\begin{thm}\label{clt-final}
When $N=1$, under assumptions $(i)$, $(ii)$, $(iii)$ and $(iv)$, we have  
\begin{equation*}
  t^{\phi^*/2}
\begin{pmatrix}
  A_t-\widetilde{A}_\infty\\
  B_t-\widetilde{A}_\infty/\phi^*
\end{pmatrix}
\stackrel{stably}\longrightarrow 
{\mathcal N}\left(0, 
\begin{pmatrix}
\phi^* \Sigma_\infty & \Sigma_\infty\\
\Sigma_\infty &\Sigma_\infty/\phi^*
\end{pmatrix}
\right)\,.
\end{equation*}
When $N\geq 2$, under assumptions $(i)$, $(ii)$, $(iii)$, $(iv)$ and $(v)$, we have 
\begin{equation*}
t^{\phi^*/2}
\begin{pmatrix}
\vA_t-\widetilde{A}_\infty\vu\\
\vB_t-\widetilde{A}_\infty/\phi^*\vu
\end{pmatrix}
\stackrel{stably}\longrightarrow 
{\mathcal N}\left(\vzero, 
\vv^\top\Sigma_\infty\vv
\begin{pmatrix}
\phi^*\vu\vu^\top & \vu\vu^\top\\
\vu\vu^\top & \vu\vu^\top/\phi^*
\end{pmatrix}
+
{\mathcal M}_\infty
\right)\,,
\end{equation*}
where ${\mathcal M}_\infty$ is given in \eqref{eq-limit-matrix}.
\end{thm}

\begin{proof}
When $N=1$, we simply have 
$(A_t,B_t)=(\widetilde{A}_t, \widetilde{A}_t/\phi^*+o(t^{-\phi^*/2}))$
 and so the statement follows from Theorem~\ref{thm:TCL_A_tilde} and Theorem~\ref{thm:CLT_N=1} for $\widetilde{A}_t$.
When $N\geq 2$, we have 
\begin{equation*}
\begin{split}
&t^{\phi^*/2}
(\vA_t-\widetilde{A}_\infty\vu, \vB_t-\widetilde{A}_\infty/\phi^*\vu)\\
&=t^{\phi^*/2}( (\widetilde{A}_t-\widetilde{A}_\infty)\vu + \widehat{\vA}_t, 
(\widetilde{A}_t-\widetilde{A}_\infty)/\phi^*\vu + \widehat{\vB}_t)\\
&=t^{\phi^*/2}(\widetilde{A}_t-\widetilde{A}_\infty) (\vu, \vu/\phi^*) + 
t^{\phi^*/2}(\widehat{\vA}_t, \widehat{\vB}_t)\,.
\end{split}
\end{equation*}
The first term converges in the sense of the almost sure conditional
convergence toward the Gaussian kernel 
with mean $\vzero$ and covariance matrix
$$
\vv^\top\Sigma_\infty\vv
\begin{pmatrix}
\phi^*\vu\vu^\top & \vu\vu^\top\\
\vu\vu^\top & \vu\vu^\top/\phi^*
\end{pmatrix}\,.
$$
The second term converges stably toward the Gaussian kernel ${\mathcal N}(\vzero, {\mathcal M}_\infty)$. 
By Lemma \ref{blocco}, we can combine the two convergences and obtain the desidered convergence. 
\end{proof}

As an immediate corollary, we can get the
central limit theorem for $\widetilde{B}_t=\vv^\top\vB_t$ when $N\geq 2$.

\begin{cor}[{CLT for $(\widetilde{B}_t)$ when $N\geq 2$}]\label{thm:TCL_B_tilde}
When $N\geq 2$, under assumptions $(i)$, $(ii)$, $(iii)$, $(iv)$ and $(v)$, we have
$$
t^{\phi^{*}/2}(\widetilde{B}_t-\widetilde{A}_\infty/\phi^*){\stackrel{stably}\longrightarrow}
\mathcal{N}\left(0\ , \vv^\top{\Sigma}_\infty\vv/\phi^*\right)\,.
$$
\end{cor}

\begin{proof}
Setting $\widetilde{B}_t=\vv^{\top}\vB_t$ and using the relations
$\vv^{\top}\vu=1$ and
$\vv^{\top}U=\vzero$ in
 Theorem~\ref{clt-final}, we find the desidered convergence.
 \end{proof}
 
We can also derive the following result that consists
  in a CLT for a normalized version of $(\vA_t,\vB_t)$.
  \begin{cor}[{CLT for $(\vA_t/(\vx^\top \vA_t), \vB_t/(\vx^\top \vB_t))$ when $N\geq 2$}]\label{thm:TCL_ratio}
    Under assumptions
    $(i)$, $(ii)$, $(iii)$, $(iv)$ and $(v)$,
    if $\widetilde{A}_\infty\neq 0$ a.s., then,
    for any given vector with $\vx^\top\vu\neq 0$, we have 
\begin{equation*}
t^{\phi^*/2}
\begin{pmatrix}
\vA_t/(\vx^\top \vA_t)-\vu/(\vx^\top \vu)\\
\vB_t/(\vx^\top \vB_t)-\vu/(\vx^\top \vu)
\end{pmatrix}
\stackrel{stably}\longrightarrow 
         {\mathcal N}\left(\vzero, \frac{1}{(\vx^\top\vu)^2\widetilde{A}_\infty^2}
         \begin{pmatrix}
 Q_\vx \mathcal{M}_\infty^{11} Q_\vx^\top & \phi^*  Q_\vx \mathcal{M}_\infty^{13} Q_\vx^\top\\
\phi^* Q_\vx \mathcal{M}_\infty^{13} Q_\vx^\top &(\phi^*)^2  Q_\vx \mathcal{M}_\infty^{33}
Q_\vx^\top
         \end{pmatrix}
\right),
\end{equation*}
where $ Q_\vx=I-\vu\vx^\top/(\vx^\top \vu)$ and the 
matrices $\mathcal{M}_{\infty}^{11},\,
\mathcal{M}_{\infty}^{13},\,\mathcal{M}_{\infty}^{33}$ are defined
in~\eqref{eq-limit-matrix}. In particular, for $\vx=\vv$ we have
$$
t^{\phi^*/2}
\begin{pmatrix}
\vA_t/\widetilde{A}_t-\vu\\
\vB_t/\widetilde{B}_t-\vu
\end{pmatrix}
\stackrel{stably}\longrightarrow 
{\mathcal N}\left(\vzero, \frac{1}{\widetilde{A}_\infty^2}
\begin{pmatrix}
\mathcal{M}_\infty^{11} & \phi^* \mathcal{M}_\infty^{13}\\
\phi^* \mathcal{M}_\infty^{13} &(\phi^*)^2 \mathcal{M}_\infty^{33}
\end{pmatrix}
\right)\,.
$$
  \end{cor}

  \begin{rem}\label{rem:TCL_ratio}
    Note that, by \eqref{definitions-hat} and
  \eqref{matrices-prop}, we have
  $\widehat{\vA}_t=\vA_t-\widetilde{A}_t\vu=UV^\top\vA_t$ and, similarly,
  $\vB_t-\widetilde{B}_t\vu=UV^\top\vB_t$. Therefore, the processes
  in the above result in the case $\vx=\vv$ coincide, up to
  multiplicative (random) factors, with
  the processes $UV^\top\vA_t$ and $UV^\top\vB_t$ (note also that
  we have $Q_{\vv}=I-\vu\vv^\top/(\vv^\top \vu)=UV^\top$ by
  \eqref{matrices-prop}).
  These processes will play a fundamental role in the following section
  dedicated to the statistical tools.
\end{rem}

\begin{proof}
The result follows from the CLT of Theorem~\ref{clt-final} once we note that 
$$\begin{aligned}
t^{\phi^*/2}&
\begin{pmatrix}
\vA_t/(\vx^\top \vA_t)-\vu/(\vx^\top \vu)\\
\vB_t/(\vx^\top \vB_t)-\vu/(\vx^\top \vu)
\end{pmatrix} \\
&=
t^{\phi^*/2}
\begin{pmatrix}
(\vx^\top \vA_t)^{-1}I & 0 \\
0 & (\vx^\top \vB_t)^{-1}I\\
\end{pmatrix}
\begin{pmatrix}
\vA_t-\vu\vx^\top \vA_t/(\vx^\top \vu)\\
\vB_t-\vu\vx^\top \vB_t/(\vx^\top \vu)
\end{pmatrix}\\
&=
t^{\phi^*/2}
\begin{pmatrix} 
(\vx^\top \vA_t)^{-1}Q_\vx & 0 \\
0 & (\vx^\top \vB_t)^{-1}Q_\vx\\
\end{pmatrix}
\begin{pmatrix}
\vA_t\\
\vB_t
\end{pmatrix}\\
&=
t^{\phi^*/2}
\begin{pmatrix} 
(\vx^\top \vA_t)^{-1}Q_\vx & 0 \\
0 & (\vx^\top \vB_t)^{-1}Q_\vx\\
\end{pmatrix}
\begin{pmatrix}
\vA_t-\widetilde{A}_\infty\vu\\
\vB_t-\widetilde{A}_\infty/\phi^*\vu
\end{pmatrix}.
\end{aligned}
$$
Indeed, we have $Q_\vx\vu\vu^\top=0$ and so, since $\vx^\top \vA_t \to (\vx^\top\vu)\widetilde{A}_\infty$ and $\vx^\top \vB_t \to (\vx^\top\vu)\widetilde{A}_\infty/\phi^*$, we have
\begin{equation*}
  \begin{split}
&\begin{pmatrix} 
(\vx^\top \vA_t)^{-1}Q_\vx & 0 \\
0 & (\vx^\top \vB_t)^{-1}Q_\vx\\
\end{pmatrix}
\left(
\vv^\top\Sigma_\infty\vv
\begin{pmatrix}
\phi^*\vu\vu^\top & \vu\vu^\top\\
\vu\vu^\top & \vu\vu^\top/\phi^*
\end{pmatrix}
+
{\mathcal M}_\infty
\right)
\begin{pmatrix} 
(\vx^\top \vA_t)^{-1}Q_\vx & 0 \\
0 & (\vx^\top \vB_t)^{-1}Q_\vx\\
\end{pmatrix}^\top
=
\\
&
\begin{pmatrix} 
(\vx^\top \vA_t)^{-1}Q_\vx & 0 \\
0 & (\vx^\top \vB_t)^{-1}Q_\vx\\
\end{pmatrix}
    {\mathcal M}_\infty
\begin{pmatrix} 
(\vx^\top \vA_t)^{-1}Q_\vx^\top & 0 \\
0 & (\vx^\top \vB_t)^{-1}Q_\vx^\top\\
\end{pmatrix}
\stackrel{a.s.}\longrightarrow
\\
&\frac{1}{(\vx^\top\vu)^2\widetilde{A}_\infty^2}
         \begin{pmatrix}
 Q_\vx \mathcal{M}_\infty^{11} Q_\vx^\top & \phi^*  Q_\vx \mathcal{M}_\infty^{13} Q_\vx^\top\\
\phi^* Q_\vx \mathcal{M}_\infty^{13} Q_\vx^\top &(\phi^*)^2  Q_\vx \mathcal{M}_\infty^{33}
Q_\vx^\top
         \end{pmatrix}
\,.
  \end{split}
  \end{equation*}
The result for $\vx=\vv$ simply follows
by the fact that, since \eqref{matrices-prop}, we have the equalities 
$Q_\vx=Q_{\vv}=(I-\vu\vv^\top/(\vv^\top \vu))=UV^\top$ and
so $UV^\top{\mathcal M}_\infty^{ij}VU^\top={\mathcal M}_\infty^{ij}$.
\end{proof}


\section{Statistical tools}
\label{app:statistical_tools}

In this section we present some statistical tools, based on Theorem~\ref{clt-final},
to provide an interval estimation of the random limit $\widetilde{A}_\infty$ and an hypothesis test on the structure of the matrix $\Phi$.
To this end, we need to assume that the 
asymptotic conditional covariance matrix $\Sigma_\infty$ of the martingale vector $\Delta\vM_t$ satisfies the following assumption:
\begin{asmp}\label{ass-matrix-Sigma}
Assume the matrix $\Sigma_\infty$ can be factorized as
\begin{equation*}
  \Sigma_\infty=g(\widetilde{A}_\infty)\Sigma_{det}\,,
  \end{equation*}
with  $g$ being a continuous function with
  $g(\widetilde{A}_\infty)>0$ a.s. and
$\Sigma_{det}$ a deterministic positive definite symmetric 
matrix determined by the eigen-structure of $\Phi$.
\end{asmp}

We will provide statistical tools based on the process $(\vA_t)$ or
on the process $(\vB_t)$ so that
 the suitable tool can be selected in the practical applications when only one of the two processes is observable.

\subsection{Confidence intervals for the random limit $\widetilde{A}_\infty$ based on $(\vA_t)$}

First of all, we can observe that by means of the central limit theorem
for the stochastic process $\widetilde{A}_{t}={\vv}^{\top}\,{\vA}_{t}$ described in Theorem~\ref{thm:TCL_A_tilde}, it is possible to
construct an asymptotic confidence
interval for its limit $\widetilde{A}_\infty$. Specifically, an asymptotic confidence interval for
$\widetilde{A}_\infty$ with approximate level $(1-\alpha)$ is the following:
\begin{equation*}
CI_{1-\alpha}(\widetilde{A}_\infty)\ :=\
\left(\
\widetilde{A}_{t}-\frac{z_{\alpha}}{t^{\phi^{*}/2}}
\sqrt{g(\widetilde{A}_t)\phi^*\vv^\top\Sigma_{det}\vv}\ ;\
\widetilde{A}_{t}+\frac{z_{\alpha}}{t^{\phi^{*}/2}}
\sqrt{g(\widetilde{A}_t)\phi^*\vv^\top\Sigma_{det}\vv}\ \right),
\end{equation*}
where $z_\alpha$ is such that
${\mathcal N}(0,1)(z_\alpha,+\infty)=\alpha/2$. 
  Note that the term $g(\widetilde{A}_\infty)$ in the confidence interval
  has been replaced by its strongly consistent estimator $g(\widetilde{A}_{t})$. \\
\indent Since the convergence in Theorem~\ref{thm:TCL_A_tilde} is also in the sense of the almost sure conditional convergence with respect to~$({\mathcal F}_t)$,  
the above interval can be recovered as a credible interval (see~\cite{Fortini20}) in a Bayesian framework.

\subsection{Confidence intervals for the random limit $\widetilde{A}_\infty$ based on $(\vB_t)$}

\indent We here construct an
asymptotic confidence interval for $\widetilde{A}_{\infty}$ based on the process $(\vB_t)_t$. 
First notice that using Theorem~\ref{clt-final} we can obtain for $N=1$
\begin{equation*}
CI_{1-\alpha}(\widetilde{A}_\infty)\ =\
\phi^{*}B_t\ \pm\ \frac{z_{\alpha}}{t^{\phi^{*}/2}}
\sqrt{g(\phi^{*}B_t )\phi^*\Sigma_{det}},
\end{equation*}
where $g(\phi^{*}B_t )$ represents a consistent estimator of $g(\widetilde{A}_\infty)$ and $z_\alpha$ is such that
${\mathcal N}(0,1)(z_\alpha,+\infty)=\alpha/2$.

\indent When $N\geq 2$, it is required the knowledge of ${\mathcal Re}(\phi_2^*)$.
Indeed,  when ${\mathcal Re}(\phi_2^*)/\phi^{*}<1/2$, setting
$\widetilde{B}_t={\vv}^{\top}{\vB}_t$ and using Corollary \ref{thm:TCL_B_tilde},
we obtain 
\begin{equation*}
CI_{1-\alpha}(\widetilde{A}_\infty)\ =\
\phi^*\widetilde{B}_t\ \pm\ \frac{z_{\alpha}}{t^{\phi^{*}/2}}
\!\sqrt{g(\phi^{*}\widetilde{B}_t )
\phi^*\vv^\top\Sigma_{det}\vv},
\end{equation*}
where $z_\alpha$ is such that
${\mathcal N}(0,1)(z_\alpha,+\infty)=\alpha/2$.

\subsection{Hypothesis tests on the matrix $\Phi$ based on $(\vA_t)$}
\label{subsec-test-A}

\indent We now focus on the case $N\geq 2$ and
the inferential problem of testing the null 
hypothesis  $H_0:\,\Phi=\Phi_0$, using the
multi-dimensional stochastic process $(\vA_t)_t$. Since the distribution of $\widetilde{A}_\infty$ should be unknown in practice,
we propose a test statistics whose limit does not involve
$\widetilde{A}_\infty$. First, we need to introduce some notation.
Let us define the matrix ${\mathcal{M}}_{det}^{11}$ as the matrix 
$\mathcal{M}_{\infty}^{11}$ defined in Theorem~\ref{clt-hat-N} with 
$\Sigma_\infty$ replaced by $\Sigma_{det}$, i.e.
$\mathcal{M}_{\infty}^{11}=
g(\widetilde{A}_\infty){\mathcal{M}}_{det}^{11}$.
Then, notice that the matrix ${\mathcal{M}}_{det}^{11}$ has rank
$(N-1)$ and  it admits a spectral decomposition
as follows: ${\mathcal{M}}_{det}^{11}=
  O^{11}{\mathcal D}^{11} (O^{11})^{\top}$,
  where ${\mathcal D}^{11}$ is the diagonal matrix of dimension $(N-1)$
  containing the eigenvalues of ${\mathcal{M}}_{det}^{11}$ different from $0$
  and $O^{11}$ is the $N\times (N-1)$ matrix whose columns 
  form an  
  orthonormal basis of $Im(\mathcal{M}_{det}^{11})$ of corresponding
  eigenvectors.   
Then, we set $L^{11}=({\mathcal D}^{11})^{-1/2}$
(diagonal matrix of dimension $(N-1)$) and take the product
$L^{11}(O^{11})^\top$ 
(matrix of dimension $(N-1)\times N$). 
Fixed the matrix assumed under $H_0$,
i.e. $\Phi_0$, we can compute for it the eigenvectors ${\vu}$ and ${\vv}$,
the eigenvectors contained in the matrices $U$ and $V$ and the matrices
$L^{11}$ and $O^{11}$ defined above.
Hence,
we can obtain under $H_0$ the real process
$\widetilde{A}_{t}={\vv}^{\top}\,{\vA}_{t}$
and the multi-dimensional process
$\widehat{{\vA}}_{t}=\vA_t-\widetilde{A}_t\vu= 
U\,V^{\top}\,{\vA}_{t}$. 
Then, from Theorem~\ref{clt-hat-N} we have that, under $H_0$
and the assumption ${\mathcal Re}(\phi_2^*)/\phi^{*}<1/2$,
\begin{equation*}\label{def:test_statistics_a}
{\vT}_{t}= t^{\phi^{*}/2}
g(\widetilde{A}_t)^{-1/2}
L^{11}(O^{11})^\top\,U\,V^{\top}\,{\vA}_{t}= t^{\phi^{*}/2}
g(\widetilde{A}_t)^{-1/2}
L^{11}(O^{11})^\top\,\widehat{{\vA}}_{t} 
\stackrel{d}\longrightarrow \mathcal{N}\left(\vzero, I_{N-1}\right)
\end{equation*}
(in the above formulas the
matrices $L^{11}$ and $O^{11}$ are related to
${\mathcal{M}}_{det}^{11}$ 
computed under $H_0$ and so for $\Phi_0$).  Hence, under $H_0$,
the test statistics
$\|{\vT}_{t}\|^2$ is 
asymptotically chi-squared distributed with $(N-1)$ degrees of
freedom. This result lets us construct an asymptotic critical region
for testing any $\Phi_0$.

\begin{rem}\label{rem-power} {\em (Power)}\\
 The performance in terms of power of this
 inferential procedure is strongly related to the matrix $\Phi_1$
 considered under the alternative hypothesis $H_1$.
 For instance, the leading eigenvalue of $\Phi$ under $H_0$ may not be equal
 to the one under $H_1$ and so the test statistics may go
 to infinity with $t$ at different rates under $H_0$ or $H_1$.
 Moreover, the leading right eigenvector ${\vv}_0$ computed under $H_0:\,
 \Phi=\Phi_0$ may not be equal to the leading right eigenvector
 ${\vv}_1$ of $\Phi_1$ and so it is possible that
$({\vv}_0^{\top}{\vA}_{t})\neq ({\vv}_1^{\top}{\vA}_{t})$.
 In that case, we would have, under $H_1$,
 that ${\vA}_{t}\stackrel{a.s.}\longrightarrow\widetilde{A}_{\infty}{\vu}_1$ but
$({\vv}_0^{\top}{\vA}_{t})\stackrel{a.s.}\longrightarrow
({\vv}_0^{\top}{\vu}_1)\widetilde{A}_{\infty}$ and
 hence, since we use $\widetilde{A}_{t}={\vv}_0^{\top}{\vA}_{t}$
 in the test statistics, 
we can be sure that
$g(\widetilde{A}_t)\stackrel{a.s.}
\longrightarrow g(\widetilde{A}_\infty)$ remains valid under
$H_1$ only if $({\vv}_0^{\top}{\vu}_1)=1$.
This condition is guaranteed in a framework where the leading left
eigenvector $\vu$ of $\Phi$ is known in advanced
(e.g. under the popular balance condition $\Phi^\top \vone = \vone$)
and so in that case we would always have $\vu_0=\vu_1=\vu$.
Analogously, the columns of $U_1$ and $V_1$ computed
under $H_1:\, \Phi=\Phi_1$ may not be equal to the corresponding
$U_0$ and $V_0$ derived under $H_0:\, \Phi=\Phi_0$. However,
if we assume $V_0^\top\vu_1=\vzero$ (that is automatically true
when $\vu_0=\vu_1$) we can be sure that, under
$H_1$, we have $U_0V_0^{\top}{\vA}_t=U_0V_0^{\top}\widehat{{\vA}}_t$, since
$$U_0V_0^{\top}{\vA}_t=
U_0V_0^{\top}(\vA_t-\widetilde{A}_t\vu_1)=
U_0V_0^{\top}\widehat{{\vA}}_t.$$
In summary, whenever $\vu_0=\vu_1$ we have under $H_1$ that
${\vT}_{t}$ is still asymptotically normally distributed with
covariance matrix
$L^{11}(O^{11})^\top UV^{\top}{\mathcal{M}}_{det}^{11}VU^{\top}
O^{11}(L^{11})^{\top}$,
where ${\mathcal{M}}_{det}^{11}$ is referred to the
eigen-structure of $\Phi_1$, while $L^{11}(O^{11})^\top UV^{\top}$ is
referred to $\Phi_0$.
Thus, the
distance between the identity $I$ and the matrix 
$L^{11}(O^{11})^\top UV^{\top}{\mathcal{M}}_{det}^{11}VU^{\top}
O^{11}(L^{11})^\top$
describes the relation between the asymptotic distribution of
$\|{\vT}_{t}\|^2$ under $H_0$ and the one under $H_1$,
which determines the power of the test.  For instance, note that
$E[\|{\vT}_{t}\|^2]=(N-1)$ under $H_0$, while
$E[\|{\vT}_{t}\|^2]$ is equal to the trace of
$L^{11}(O^{11})^\top UV^{\top}{\mathcal{M}}_{det}^{11}VU^{\top}O^{11}(L^{11})^\top
$
under $H_1$.
\end{rem}

\subsection{Hypothesis tests on the matrix $\Phi$ based on $(\vB_t)$}
We here want to build for the case $N\geq 2$
a test statistic for $H_0: \Phi=\Phi_0$,
that is based only on the process $(\vB_{t})$.
To this end, analogously as in Subsection~\ref{subsec-test-A},
let us define the matrix
${\mathcal{M}}_{det}^{33}$ such that
$\mathcal{M}_{\infty}^{33}=
g(\widetilde{A}_\infty){\mathcal{M}}_{det}^{33}$
and the corresponding matrices $L^{33}$ and $O^{33}$. 
Fixed the matrix $\Phi_0$ assumed under $H_0$,
we can obtain the real process
$\widetilde{B}_{t}={\vv}^{\top}\,{\vB}_{t}$
and the multi-dimensional process
$$\vB'_t=\vB_t-\widetilde{B}_t\vu=U\,V^{\top}\,{\vB}_{t}.
$$
By Corollary~\ref{thm:TCL_ratio} and Remark~\ref{rem:TCL_ratio},
we get that,
under $H_0$ and the assumption ${\mathcal Re}(\phi_2^*)/\phi^{*}<1/2$, we have 
\begin{equation*}\label{def:test_statistics_b}
{\vT}_{t}= t^{\phi^{*}/2}
g(\phi^{*}\widetilde{B}_t)^{-1/2}L^{33}(O^{33})^\top\vB'_t =
t^{\phi^{*}/2}
g(\phi^{*}\widetilde{B}_t)^{-1/2}
L^{33}(O^{33})^\top\,U\,V^{\top}\,{\vB}_{t}
\stackrel{d}\longrightarrow\mathcal{N}\left(\vzero,I_{N-1}\right)
\end{equation*}
(in the above formulas the matrices $L^{33}$ and $O^{33}$ are  related to
${\mathcal{M}}_{det}^{33}$
computed under $H_0$ and so for $\Phi_0$). 
The considerations on the power made for the test based on $\vA_t$
in Remark~\ref{rem-power} remain
valid as well for the test based on $\vB_t$. 


\section{Statistical tools in some examples}
\label{app:statistical_tools_examples}

We here specify the previous general statistical tools in two particular
cases: firstly, we consider the case $N=2$ and, then, we take into account
the case when $\Phi$ is of the ``mean-field'' type and $N\geq 2$.

\subsection{Case $N = 2$}

A matrix $\Phi$ with dimension $2\times 2$ and satisfying condition $(i)$
 can be reparametrized
imposing that the number $\phi\in (0,1]$ is an eigenvalue with
left eigenvector $\vu$ such that $r=u_1/u_2$, with $r>0$, that is
$$
r\phi_{11}+\phi_{21}=r\phi
\quad\mbox{and}\quad
r\phi_{12}+\phi_{22}=\phi\,.
$$
This reparametrization  has been inspired by the fact that
we typically are able to estimate  with high accuracy 
the leading eigenvalue $\phi^{*}\in (0,1]$ and
  the ratio $r=u_1/u_2>0$ of the two componets of the
  leading left eigenvector $\vu$. Hence, we may think that
  $\phi=\phi^*$ and $r$ are known in practice and the only parameters
  that are object of the test  are $\phi_{21}$ and $\phi_{12}$.
  We can also calculate the second eigenvalue from the trace of $\Phi$:
$$\phi_2^{*}=\phi_{1,1}+\phi_{2,2}-\phi=\phi-\frac{\phi_{2,1}+r^2\phi_{1,2}}{r},$$
so that the condition ${\mathcal Re}(\phi_2^*)/\phi^{*}<1/2$
required in the previous results 
corresponds to the condition
$$
2\frac{\phi_{2,1}+r^2\phi_{1,2}}{r} > \phi.
$$
  Moreover, since according to our notation
  we have $\vv^\top \vone=1$ and $\vv^\top \vu=1$, we obtain
  the following relations:
$$
v_1 = \frac{r\phi_{1,2}}{r\phi_{1,2}+\phi_{2,1}},\qquad
v_2 = 1-v_1,\qquad
u_1 = \frac{r}{1-(1-r)v_1},\qquad
u_2 = \frac{1}{1-(1-r)v_1},
$$
which leads to
$$
v_1 = \frac{r\phi_{1,2}}{r\phi_{1,2}+\phi_{2,1}},\qquad
v_2 = \frac{\phi_{2,1}}{r\phi_{1,2}+\phi_{2,1}},\qquad
u_1 = \frac{r\phi_{2,1}+r^2\phi_{1,2}}{
\phi_{2,1}+r^2\phi_{1,2}},\qquad
u_2 = \frac{\phi_{2,1}+r\phi_{1,2}}{
\phi_{2,1}+r^2\phi_{1,2}}.
$$
If we adopt the parametrization  $\iota=\phi_{1,2}+\phi_{2,1}$ and $\eta=\tfrac{\phi_{1,2}}{\phi_{1,2}+\phi_{2,1}}$
so that $\iota$ represents the intensity of the whole ``interaction'' among the two components $1$ and $2$ due to matrix $\Phi$ 
and $\eta$ (resp. $(1-\eta)$) gives the percentage of this intensity due to the influence 
of the first component on the second one (resp., of the second component on the first one). 
Then, we have
$$
\vu = \frac{(1-\eta)+r\eta}{
(1-\eta)+r^2\eta}
\begin{pmatrix}
r\\
1
\end{pmatrix}
\qquad\mbox{and}\qquad
\vv = \frac{1}{r\eta+(1-\eta)}
\begin{pmatrix}
r\eta\\
1-\eta
\end{pmatrix},
$$
and so
$$
\vu\vv^\top=
\frac{1}{(1-\eta)+r^2\eta}
\begin{pmatrix}
r\\
1
\end{pmatrix}
\begin{pmatrix}
r\eta &
1-\eta
\end{pmatrix}
\qquad\mbox{and}\qquad
UV^\top=
\frac{1}{(1-\eta)+r^2\eta}
\begin{pmatrix}
1-\eta\\
-r\eta
\end{pmatrix}
\begin{pmatrix}
1 &
-r
\end{pmatrix}.
$$
Finally, imposing $U^\top U=1$, we get 
$$
U = \frac{1}{\sqrt{(1-\eta)^2+r^2\eta^2}}
\begin{pmatrix}
1-\eta\\
-r\eta
\end{pmatrix}
\qquad\mbox{and}\qquad
V=\frac{\sqrt{(1-\eta)^2+r^2\eta^2}}{(1-\eta)+r^2\eta}
\begin{pmatrix}
1 \\
-r
\end{pmatrix}.
$$
We thus
obtain 
$$\phi^*\vv^\top\Sigma_{det}\vv=
\frac{\phi}{(r\eta+(1-\eta))^2}
\begin{pmatrix}
r\eta & 1-\eta
\end{pmatrix} 
\Sigma_{det}
\begin{pmatrix}
r\eta \\ 1-\eta 
\end{pmatrix} $$
and, for $2\iota\frac{(1-\eta)+r^2\eta}{r} > \phi$,
$$
{\mathcal{M}}_{det}^{11}=U{\mathcal{S}}_{det}^{11}U^\top = 
\frac{(\iota\frac{(1-\eta)+r^2\eta}{r}-\phi)^2}{2\iota\frac{(1-\eta)+r^2\eta}{r}-\phi}
\frac{1}{((1-\eta)+r^2\eta)^2}
\begin{pmatrix}
1-\eta\\
-r\eta
\end{pmatrix}
\begin{pmatrix}
1 &
-r
\end{pmatrix}
\Sigma_{det}
\begin{pmatrix}
1\\
-r
\end{pmatrix}
\begin{pmatrix}
1-\eta &
-r\eta
\end{pmatrix},
$$
$$
{\mathcal{M}}_{det}^{33}=
U{\mathcal{S}}_{det}^{33}U^\top = 
\frac{1}{2\iota\frac{(1-\eta)+r^2\eta}{r}-\phi}
\frac{1}{((1-\eta)+r^2\eta)^2}
\begin{pmatrix}
1-\eta\\
-r\eta
\end{pmatrix}
\begin{pmatrix}
1 &
-r
\end{pmatrix}
\Sigma_{det}
\begin{pmatrix}
1\\
-r
\end{pmatrix}
\begin{pmatrix}
1-\eta &
-r\eta
\end{pmatrix}.
$$
(where $\mathcal{S}_{det}^{ii}$ are
  the matrices defined in Theorem~\ref{clt-hat-N} with 
  $\Sigma_\infty$ replaced by $\Sigma_{det}$).
Therefore, recalling that 
$$\widetilde{A}_t=\vv^\top \vA_t=\frac{r\eta A_{1,t}+(1-\eta)A_{2,t}}{r\eta+(1-\eta)}
\qquad\mbox{and}\qquad
\widetilde{B}_t=\vv^\top \vB_t=\frac{r\eta B_{1,t}+(1-\eta)B_{2,t}}{r\eta+(1-\eta)},
$$ and defining the constant
$$
c_{r,\eta} =   \tfrac{\begin{pmatrix}
r\eta & (1-\eta)
\end{pmatrix} 
\Sigma_{det}
\begin{pmatrix}
r\eta \\ (1-\eta)
\end{pmatrix}}{(r\eta+(1-\eta))^2},
$$
the confidence interval based on $\vA_t$ and $\vB_t$ are respectively 
$$
CI_{1-\alpha}(\widetilde{A}_\infty)\ =\
\widetilde{A}_t \pm\ \frac{z_{\alpha}}{t^{\phi/2}}
\!\sqrt{ \phi g(\widetilde{A}_t)c_{r,\eta}}\,,
\qquad\mbox{and}\qquad
CI_{1-\alpha}(\widetilde{A}_\infty)\ =\
\phi\widetilde{B}_t \pm\ \frac{z_{\alpha}}{t^{\phi/2}}
\!\sqrt{ \phi g(\phi\widetilde{B}_t)c_{r,\eta}}.
$$
Analogously, the test statistics based on $\vA_t$ is 
$$
T_{t}= t^{\phi/2}
\frac{1}{(1-\eta)+r^2\eta}
L^{11}(O^{11})^\top \begin{pmatrix}
(1-\eta)\\
-r\eta
\end{pmatrix}
\frac{(A_{1,t}-r A_{2,t})}{g(\widetilde{A}_t)^{1/2}},
$$
and the one based on $\vB_t$
$$
T_{t}= t^{\phi/2}
\frac{1}{(1-\eta)+r^2\eta}
L^{33}(O^{33})^\top \begin{pmatrix}
(1-\eta)\\
-r\eta
\end{pmatrix}\,
\frac{(B_{1,t}-r B_{2,t})}{g(\phi\widetilde{B}_t)^{1/2}}.
$$

\begin{rem}\label{rem:N_equal_2_Sigma_diagonal}
In the particular case $\Sigma_{det}=diag(\vu)$ we have 
$$
{\mathcal{S}}_{det}^{33} =
\frac{1}{\left(2\iota \frac{(1-\eta)+r^2\eta}{r}-\phi\right)}
\frac{(r+r^2)((1-\eta)+r\eta)((1-\eta)^2+r^2\eta^2)}{((1-\eta)+r^2\eta)^3},
$$
which leads to $L^{33}(O^{33})^\top=({\mathcal{S}}_{det}^{33})^{-1/2}U^\top$
and so the test statistics based on $\vB_t$ becomes
$$
T_{t}= t^{\phi/2}
\sqrt{\left(2\iota \frac{(1-\eta)+r^2\eta}{r}-\phi\right)\frac{(1-\eta)+r^2\eta}{(r+r^2)((1-\eta)+r\eta)}}
\,\frac{(B_{1,t}-rB_{2,t})}{g(\phi\widetilde{B}_t)^{1/2}}.
$$
Then, denoting
$$
\Delta =\frac{1}{2}-\frac{\phi^{*}_2}{\phi}= 
\frac{\iota}{\phi}\frac{(1-\eta)+r^2\eta}{r}-\frac{1}{2}
\qquad\mbox{and}\qquad
q_{r,\eta} = \frac{2((1-\eta)+r^2\eta)}
{r(1+r)((1-\eta)+r\eta)},
$$
we have 
\begin{equation}\label{test}
  \|T_{t}\|^2= t^{\phi}q_{r,\eta}\Delta
\,\frac{(B_{1,t}-rB_{2,t})^2}{g(\phi \widetilde{B}_t)/\phi}
\stackrel{d}\longrightarrow\chi^2(N-1).
\end{equation}
Moreover, the confidence interval based on $\vB_t$ becomes
\begin{equation}\label{conf-interval}
CI_{1-\alpha}(\widetilde{A}_\infty)\ =\
\phi\widetilde{B}_t \pm\ \frac{z_{\alpha}}{t^{\phi/2}}
\!\sqrt{ \phi g( \phi\widetilde{B}_t)c_{r,\eta}}\,,
\qquad\mbox{with}\qquad
c_{r,\eta} =   \tfrac{(r^3\eta^2+(1-\eta)^2)}{(r\eta+(1-\eta))(r^2\eta+(1-\eta))}.
\end{equation}
\end{rem}


\subsection{Mean-field case for $N\geq 2$}\label{subsection_medione}

This first case we consider is the ``Mean-field'' case, which refers to the family of matrices $\Phi$
that can be expressed as follows:
for any $1\leq h, j\leq N$
\begin{equation*}
\phi_{j,h}\ =\ \phi\left(\frac{\iota}{N}\ +\ \delta_{j,h}(1-\iota)\right)
\qquad\mbox{with } \phi,\iota\in (0,1],
\end{equation*}
where $\delta_{j,h}$ is equal to $1$ when $h=j$ and to $0$
otherwise. Note that $\Phi$ is irreducible and
doubly stochastic, we have ${\vv}=N^{-1}{\vone}$ and ${\vu}=\vone$ and so the random variable
$\widetilde{A}_{t}$ coincides with the average of the processes
$A_{t,h}$, i.e. $N^{-1}\vone^{\top}{\vA}_{t}$ and
$\widehat{{\vA}}_{t}=
\left(I-N^{-1}\vone\vone^{\top}\right){\vA}_t$.
Furthermore, in this case, the leading eigenvalue $\phi^*$ coincides with   
$\phi$, while 
all the eigenvalues of $\Phi$ different from $\phi^{*}$
are equal to $\phi(1-\iota)$ and, 
consequently, the condition
${\mathcal Re}(\phi_2^*)/\phi^{*}<1/2$
required in the previous results
corresponds to the condition $2\iota>1$.
Finally, since $\Phi$ is also symmetric, we have $U=V$ and so
$U^{\top}U=V^{\top}V=I$ and
$UU^{\top}=VV^{\top}=I-N^{-1}\vone\vone^{\top}$. Notice that the
mean-field matrix
has two parameters, $\phi$ and $\iota$, but, since the leading eigenvalue $\phi^{*}$ is typically easy to be estimated with high accuracy,
we may think that the only parameter that is object of inference in this case is $\iota$, which is the quantity that rules the intensity of the ``interaction'' among the different components due to  matrix $\Phi$.
We thus obtain 
$\phi^*\vv^\top\Sigma_{det}\vv=
\phi N^{-2}
(\vone^\top \Sigma_{det}\vone)$
and, for $2\iota>1$,
$$
{\mathcal{M}}_{det}^{11}=U{\mathcal{S}}_{det}^{11}U^\top = 
\frac{\phi(1-\iota)^2}{2\iota-1}
UV^\top \Sigma_{det} VU^\top
=
\frac{\phi(1-\iota)^2}{2\iota-1}
(I-N^{-1}\vone\vone^\top)\Sigma_{det}(I-N^{-1}\vone\vone^\top),$$
$$
{\mathcal{M}}_{det}^{33}=
U{\mathcal{S}}_{det}^{33}U^\top = 
\frac{1}{\phi(2\iota-1)}
UV^\top \Sigma_{det} VU^\top =
\frac{1}{\phi(2\iota-1)}
(I-N^{-1}\vone\vone^\top)\Sigma_{det}(I-N^{-1}\vone\vone^\top),
$$
Therefore, recalling 
$\widetilde{A}_t=\tfrac{\vone^\top \vA_t}{N}$,
$\widetilde{B}_t=\tfrac{\vone^\top \vB_t}{N}$ and defining $c_N=\tfrac{\left(\vone^\top \Sigma_{det}\vone\right)}{N^2}$,
the confidence interval based on $\vA_t$ and $\vB_t$ are respectively 
\begin{equation*}
CI_{1-\alpha}(\widetilde{A}_\infty)\ =\
\widetilde{A}_t\ \pm\ \frac{z_{\alpha}}{t^{\phi/2}}
\!\sqrt{\phi g(\widetilde{A}_t)c_N}
\qquad\mbox{and}\qquad
CI_{1-\alpha}(\widetilde{A}_\infty)\ =\
\phi\widetilde{B}_t\ \pm\ \frac{z_{\alpha}}{t^{\phi/2}}
\!
\sqrt{\phi g(\phi\widetilde{B}_t)c_N}.
\end{equation*}
Analogously,
the test statistics based on $\vA_t$ is 
$$
{\vT}_{t}=  t^{\phi/2}
L^{11}(O^{11})^\top UV^\top\,\frac{{\vA}_{t}}{g(\widetilde{A}_t)^{1/2}}=
t^{\phi/2}
L^{11}(O^{11})^\top (I-N^{-1}\vone\vone^\top)\,
\frac{{\vA}_{t}}{g(\widetilde{A}_t)^{1/2}},
$$
and the one based on $\vB_t$ is 
$$
{\vT}_{t}= 
t^{\phi/2}
L^{33}(O^{33})^\top UV^\top\,
\frac{{\vB}_{t}}{g(\phi\widetilde{B}_t)^{1/2}}
= t^{\phi/2}
L^{33}(O^{33})^\top (I-N^{-1}\vone\vone^\top)\,
\frac{{\vB}_{t}}{g(\phi\widetilde{B}_t)^{1/2}}.
$$

\begin{rem}\label{rem:mean_field_Sigma_diagonal}
In the particular case $\Sigma_{det}=I$ we have 
$\vone^\top\Sigma_{det}\vone=
N$ and hence the following confidence interval based on $\vB_t$:
\begin{equation}\label{conf-interval-mean}
CI_{1-\alpha}(\widetilde{A}_\infty)\ =\
\phi\widetilde{B}_t\ \pm\ \frac{z_{\alpha}}{t^{\phi/2}}
\!
\sqrt{\phi g(\phi\widetilde{B}_t)c_N}
\qquad\mbox{with}\qquad
c_N=\frac{1}{N}.
\end{equation}
Moreover, we have
$
{\mathcal{M}}_{det}^{33} = U{\mathcal{S}}_{det}^{33}U^\top,
$
with
$$
UU^\top=(I-N^{-1}\vone\vone^\top)
\qquad\mbox{and}\qquad
{\mathcal{S}}_{det}^{33} =
\frac{1}{\phi(2\iota-1)}I,
$$
which leads to $L^{33}(O^{33})^\top = \sqrt{\phi(2\iota-1)}U^\top$
and so the test statistics based on $\vB_t$ becomes
$$
{\vT}_{t}=t^{\phi/2}
\sqrt{\phi(2\iota-1)}U^\top\,
\frac{{\vB}_{t}}{g(\phi\widetilde{B}_t)^{1/2}},
$$
from which we can get
$$
\begin{aligned}
  \|\vT_{t}\|^2= t^\phi
 (2\iota-1)
  \frac{{\vB}_{t}^\top(I-N^{-1}\vone\vone^\top){\vB}_{t}}{g(\phi\widetilde{B}_t)/\phi}
  &=\ t^\phi
 (2\iota-1)\frac{\|\vB_{t}\|^2-\frac{(\vone^\top \vB_t)^2}{N}}{g(\widetilde{B}_t)/\phi}\\
  &=\ t^\phi 
(2\iota-1)
  \frac{\big\|\vB_t-\left(\frac{\vone^\top \vB_t}{N}\right)\vone
  \big\|^2}{g(\phi\widetilde{B}_t)/\phi},
\end{aligned}
$$
that is
\begin{equation}\label{test-mean}
\|T_{t}\|^2= t^\phi 
2\Delta
\frac{\big\|\vB_t-\widetilde{B}_t\vone
  \big\|^2}{g(\phi\widetilde{B}_t)/\phi}  
\stackrel{d}\longrightarrow \chi^2(N-1)
  \qquad\mbox{with}\qquad
\Delta =
\frac{1}{2}-\frac{\phi^{*}_2}{\phi}
=\iota-\frac{1}{2}.
\end{equation}

\end{rem}


\section{Technical results}

For the reader's convenience, we here recall some results used in the above proofs. (For a synthetic review about
 the notion of stable convergence and its variants, see for instance the appendix of \cite{ale-cri-ghi-WEIGHT-MEAN} and the references recalled therein.)

\begin{thm}[{A consequence of \cite[Theorem A.1]{crimaldi-2009}}]
\label{fam_tri_vet_as_inf}
 On~$(\Omega,\mathcal{A},P)$, let $(\mathcal{F}_{t})_t$ be a
  filtration and,  for each $t\geq 1$, let $(L_{t,k})_{k\in{\mathbb N}}$ be a real martingale with respect
  to $({\mathcal{F}}_{t+k-1})_{k}$, with $L_{t,0}=0$, which
  converges in mean to a random variable $L_{t,\infty}$. Set
\begin{equation*}
Y_{t,n}=L_{t,n}-L_{t,n-1}\quad\hbox{for } n\geq 1,\quad
U_t=\textstyle\sum_{n\geq 1} Y_{t,n}^2,\quad
Y_t^*=\textstyle\sup_{n\geq 1}\;|Y_{t,n}|.
\end{equation*}
Further, suppose that $(U_t)_t$ converges almost surely to a positive real (${\mathcal F}_\infty$-measurable) 
random variable $U$ and   
$Y_t^*=O(y_t)$ with $y_t$ a deterministic sequence such that $\lim_{t\to+\infty} y_t = 0$.
\\
\indent Then, with respect to $(\mathcal{F}_t)_t$, the sequence $(L_{t,\infty})_{t}$ converges to the
Gaussian kernel ${\mathcal{N}}(0,U)$ in the sense of the almost sure
conditional convergence (and so also stably).
\end{thm}

\begin{lem}[{\cite[Lemma~S2.2]{ale-cri-ghi-complete}}] 
\label{lemmaS2.2}  
If $a_t \geq 0$, $a_t \leq 1$ for $t$ large enough, $\sum_t a_t = +\infty$, $\delta_t \geq 0$, $\sum_t \delta_t < +\infty$, $b > 0$,
$y_t \geq 0$ and $y_{t+1} \leq (1 - a_t )^b y_t + \delta_t$, then
$\lim_{t\to+\infty} y_t = 0$.
\end{lem}

\begin{thm}[{A consequence of \cite[Proposition~3.1]{cri-pra}}]
\label{th-triangular}
Let $({\mathbf T}_{t,k+1})_{t\geq m_0, m_0\leq k\leq t}$ be a triangular
array of $d$-dimensional real random vectors, such that, for each
fixed $t$, the finite sequence $({\mathbf T}_{t,k+1})_{m_0\leq k\leq t}$
is a martingale difference array with respect to a given filtration
$({\mathcal F}_{k+1})_{k\geq m_0}$. Moreover, let $(s_t)_t$ be a
sequence of positive real numbers and assume that the following
conditions hold:
\begin{itemize}
\item[(1)] $\sum_{k=m_0}^{t} (s_t{\mathbf T}_{t,k+1})(s_t{\mathbf T}_{t,k+1})^{\top}=
s_t^2\sum_{k=m_0}^{t} {\mathbf T}_{t,k+1}{\mathbf T}_{t,k+1}^{\top} \stackrel{P}\longrightarrow {\mathcal M}_\infty$,
  where ${\mathcal M}_\infty$ is a (${\mathcal F}_\infty$-measurable) random positive semi\-defi\-ni\-te matrix;
\item[(2)] $\sup_{m_0\leq k\leq t} |s_t{\mathbf T}_{t,k+1}|
  =s_t\sup_{m_0\leq k\leq t} \sum_{i=1}^d |T_{t,k+1,i}|=O(y_t)$
  with $(y_t)$ a deterministic sequence such that $\lim_{t\to+\infty} y_t = 0$.
\end{itemize}
Then $(s_t\sum_{k=m_0}^{t}{\mathbf T}_{t,k+1})_t$ converges stably to the
Gaussian kernel ${\mathcal N}(\vzero, {\mathcal M}_\infty)$.
\end{thm}
Recall that, according to the notation used here, the above condition 2) means that 
$\sup_{m_0\leq k\leq t} |s_t{\mathbf T}_{t,k+1}|
  =s_t\sup_{m_0\leq k\leq t} \sum_{i=1}^d |T_{t,k+1,i}|\leq C y_t$ for a suitable deterministic constant $C$  and 
  $t$ large enough, so that we trivially have the convergence in $L^1$ of 
  $\sup_{m_0\leq k\leq t} |s_t{\mathbf T}_{t,k+1}|$ toward zero,  
  as requested in  \cite[Proposition~3.1]{cri-pra}.\\

The following result combines together stable convergence and
 almost sure conditional convergence. 

\begin{lem}[{A consequence of \cite[Lemma 1]{BeCrPrRi11}}]
\label{blocco}
Suppose that $(\vY^{(1)}_t)$ and $(\vY^{(2)}_t)$ are real random vectors such that 
 the first is adapted to  a filtration $({\mathcal F}_t)_t$
and the second is measurable with respect to 
${\mathcal F}_\infty$.   
If $(\vY^{(1)}_t)$ converges stably to $\mathcal{N}(0,{\mathcal M}^{(1)}_{\infty})$ and $(\vY^{(2)}_t)$  
converges to $\mathcal{N}(0,{\mathcal M}^{(2)}_{\infty})$ in the sense of
the almost sure conditional convergence
with respect to $({\mathcal F}_t)_t$, then
$$
(\vY^{(1)}_t+\vY^{(2)}_t)\stackrel{stably}\longrightarrow \mathcal{N}(\vzero,{\mathcal M}^{(1)}_\infty+{\mathcal M}^{(2)}_\infty).
$$
\end{lem}

%


\end{document}